\documentclass[a4paper,11pt,onecolumn]{article}
\usepackage{amsmath}
\usepackage{amsfonts}

\usepackage{amssymb}
\usepackage{graphicx}
\usepackage{bm}
\usepackage{geometry}
\usepackage{cite}

\usepackage[plainpages=false,hypertexnames=true,hyperindex=true,pagebackref=false,raiselinks=true]{hyperref}
\hypersetup{bookmarks,bookmarksnumbered,colorlinks,linkcolor=blue,breaklinks=true,linktoc=all}

\input{tcilatex}

\graphicspath{{./Figures/}}

\begin{document}

\title{The outcomes of measurements in the de Broglie-Bohm theory}
\author{G. Tastevin\thanks{%
tastevin@lkb.ens.fr}~ and F. Lalo\"{e}\thanks{%
laloe@lkb.ens.fr} \\
Laboratoire Kastler Brossel, ENS-Universit\'e PSL,\\
CNRS, Sorbonne Universit\'e, Coll\`ege de France,\\
24 rue Lhomond 75005\ Paris, France}
\date{\today}
\maketitle

\begin{abstract}
Within the de Broglie-Bohm (dBB) theory, the measurement process and the determination of its outcome are usually
discussed in terms of the effect of the Bohmian positions of the
measured system S. 
This article shows that the Bohmian positions associated
with the measurement apparatus M can
actually play a crucial role in the determination of the
result of measurement. Indeed, in many cases, the result is practically
independent of the initial value of a Bohmian position associated with S,
and determined only by those of M. The measurement then does not reveal
the value of any pre-existing variable attached to S, but just the initial
state of the measurement apparatus. Quantum
contextuality then appears with particular clarity as a consequence of the  dBB dynamics for entangled systems.
\end{abstract}

A well-known feature of standard quantum mechanics is that the result of a
measurement does not reveal the value of any pre-existing physical quantity
attached to the measured system S; in most situations, the result is
actually created during the interaction process between S and the
measurement apparatus. Jordan, as quoted by Bell \cite{Bertlmann}, wrote for instance:
\textquotedblleft In a measurement of position, the electron is forced .. to
assume a definite position; previously, it was neither here nor there, it
had not yet made its decision to a definite position..\textquotedblright . Within
standard quantum mechanics, quantum contextuality \cite{Bell-1966,Bell-livre,Kochen-Specker} is a well-known consequence of this property.\ In the de Broglie-Bohm (dBB)
theory \cite{De-Broglie-1927,Bohm-1952}, the result is completely determined
by the initial value of the positions, which evolve within a deterministic dynamics\cite{Albert-1992, Holland-1993, Oriols-Mompart-2012, Durr-et-al-2013, Gondran, Bricmont-2016 }. The only source of randomness arises
from the fact that these positions are unknown.\ Now, when a quantum
system S interacts with a measurement apparatus M, one may wonder
if the result is primarily determined by the Bohmian positions associated
with S, or by those associated with the macroscopic measurement apparatus M.

In most cases, the number of Bohmian positions  associated with M is much larger than
that  associated with S, and it seems natural that these positions should play a role. But, curiously, the measurement process within dBB theory is
often discussed
in a model where M is treated as a classical external
potential acting on S, which amounts to merely ignoring the quantum properties of M.
For instance, in a Stern-Gerlach experiment, the effect of the magnet on the
incoming atom is treated classically through the action of given external magnetic gradient; all
effects of quantum entanglement between S and M are then ignored.
It is then clear that the initial Bohmian position of the incoming particle
within its wave function is the only variable that can determine the final outcome (whether the
particle is deviated upwards or downwards). Indeed,  informal discussions with colleagues physicists often reveal the belief that, within the dBB theory, the result of measurement is determined by the initial value\footnote{The relation between this initial value and the result should then be contextual, as emphasized for instance in Refs~\cite{Albert-1992,Bricmont-2016,Bricmont-Goldstein-Hemmick}.} of the Bohmian variable(s) attached to S.

The role of the variables attached to M is nevertheless also discussed by some authors. For instance, Holland \cite{Holland-1993} shows how these variables introduce irreversibility in the registration process of the final result, an essential step in any real measurement process. Bricmont \cite{Bricmont-2016} gives a similar discussion in terms of effective collapse and decoherence. The possible influence on the result itself is explicitely mentioned by D\"{u}rr and Teufel \cite{Durr-Teufel}  but they consider that, in line with one's intuition, the result of measurement is often determined by the positions attached to S alone, and not by the initial value of the coordinates of the particles of the pointer particles belonging to M.

The purpose of the present article is to show that this is not the case in general, and even that the complete opposite may  be  true.
Indeed we will see that, in many cases, the measurement result is actually almost independent of the Bohmian positions attached to S; it is rather determined by the initial values of the Bohmian position of M.
The
measurement result is then a consequence of the initial physical state
of the measurement apparatus, and may appear as predetermined. This idea has some internal consistency: it seems natural that, in the interaction between a large system and a small system, the former dominates the process and forces the small system to follow a given evolution.

\section{A model of a spin measurement}
\label{model}

In the historical Stern-Gerlach experiment \cite{Stern-Gerlach}, what was
observed was the accumulation of silver atoms onto a glass slide. This method does not seem really appropriate for the measurement of single
particles. Moreover, various recoil effects affect the atoms inside the measurement apparatus: depending on the result of measurement, the atoms and molecules inside in the magnet recoil upwards or downwards, various localized phonon emission processes occur inside the glass plate, etc. Modelling these effects and the resulting changes of quantum states would be a complicated task. We will therefore introduce an optical version of the experiment
that seems to be more appropriate for a quantum treatment of single measurement events.

\subsection{Optical Stern-Gerlach experiment}

\begin{figure}[!b]  
\centering
\hrule
\includegraphics[trim = 0mm 0mm 0mm 0mm, clip,width=12cm]
{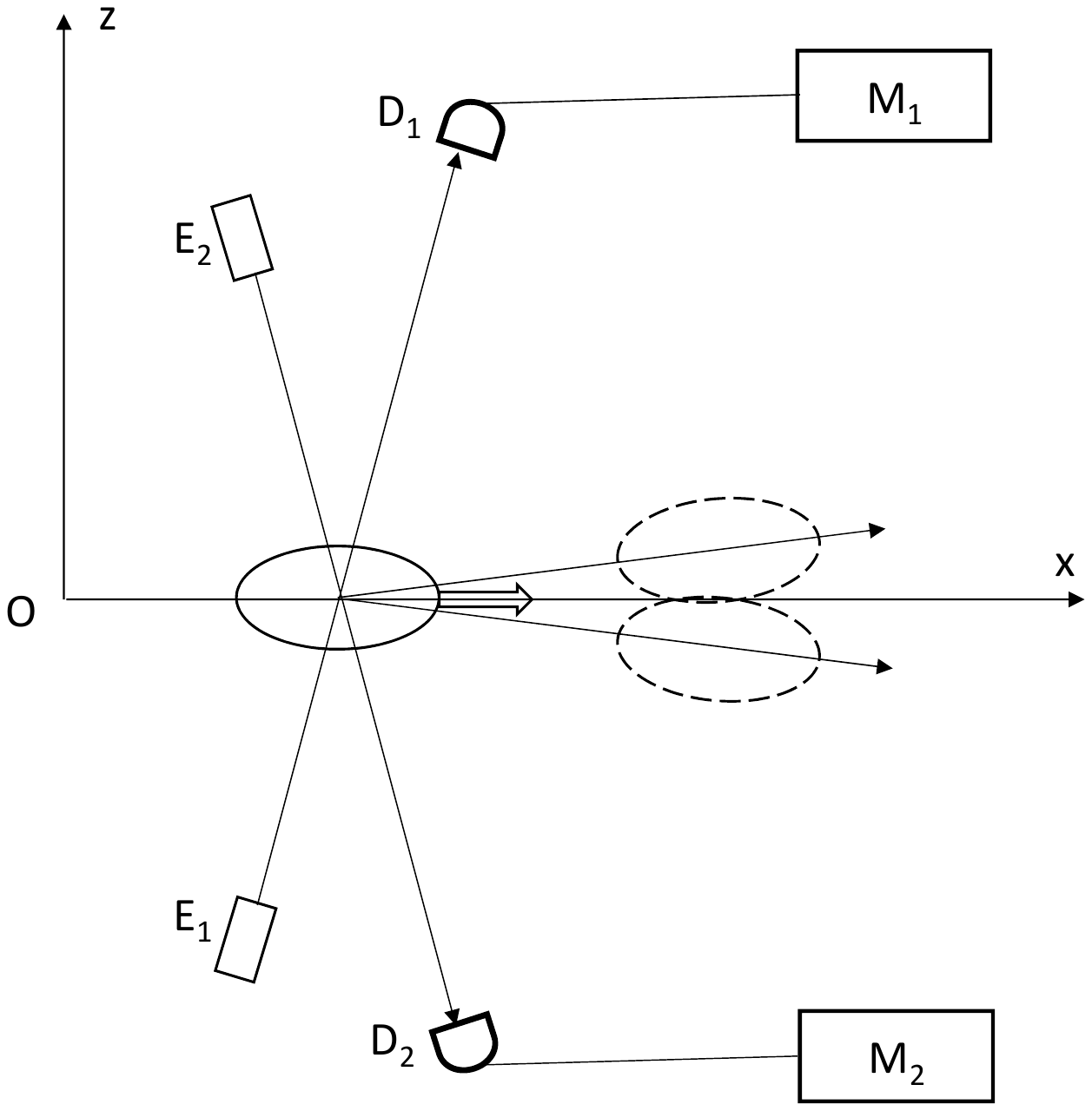}
\caption{Schematic representation of an optical version of the
Stern-Gerlach experiment. A spin $1/2$ particle is
described by a wave packet propagating in the direction of the Ox axis, and symbolized by
the solid ellipse in the figure. If the particle is in the $\left\vert +\right\rangle$ spin state, it can
absorb a circularly polarized photon emitted by E$_{1}$,
but not the photon emitted by E$_{2}$, since it has the opposite
circular polarization. The converse is true if the particle is in the spin $\left\vert -\right\rangle$
state. In the former case, the photon recoil transfers a positive momentum
to the wave packet along axis Oz, in the latter case a negative momentum.
Either photon detector D$_{2}$ or D$_{1}$ then registers a clic, and its signal
is amplified sufficiently to move the position of a macroscopic pointer displaying the
results in apparatuses M$_{2}$ or M$_{1}$. We assume that the spin particle
is initially in a coherent superposition $\protect\alpha\left\vert
+\right\rangle +\protect\beta\left\vert -\right\rangle $ of the two spin
eigenstates.}
\label{Modele-SG}
\end{figure}

The experiment is sketched in figure~\ref{Modele-SG}. The atom wave
packet propagates along the direction of the $Ox$ axis, and then
crosses two   wave packets of photons having opposite directions of propagation and circular
polarizations, emitted respectively by two single photon sources E$_1$ and E$_2$. Depending on the  spin state of the atom, one of the photons is absorbed and scattered in all directions, while the
atom undergoes a recoil that pushes it either upwards of downwards. The other non absorbed photon is detected by either detector D$_{1}$ or D$_{2}$, which
provides a measurement of the atomic spin along direction Oz. We assume
that, initially, the atom is in a coherent spin state:
\begin{equation}  \label{eq1}
\alpha\left\vert +\right\rangle +\beta\left\vert -\right\rangle
\hspace{0.5cm}
\end{equation}
with the normalization condition $\left\vert  \alpha \right\vert ^2 + \left\vert \beta \right\vert ^2 =1$.

\subsection{Wave function}

We use a notation that is similar to that of Ref. \cite{GT-FL}.\ The initial
quantum state of the whole system at time $t=0$ is symbolized by the expression:%
\begin{equation}
\left\vert \Phi(t=0)\right\rangle =\Big[ \alpha\left\vert +\right\rangle
+\beta\left\vert -\right\rangle \Big] \left\vert
\varphi_{z}^{0}\right\rangle \left\vert \varphi_{x}^{0}\right\rangle
\left\vert \varphi _{y}^{0}\right\rangle \prod \limits_{p=1}^{N}\left\vert
\chi_{1}^{0} \right\rangle _p \prod \limits_{n=1}^{N}\left\vert
\chi_{2}^{0}\right\rangle _n   \label{1}
\end{equation}
where $\left\vert \varphi_{z}^{0}\right\rangle \left\vert
\varphi_{x}^{0}\right\rangle \left\vert \varphi_{y}^{0}\right\rangle $ is the
ket describing the orbital state of the atom as a product of Oz, Ox and Oy states.
The kets $\left\vert \chi_{1}^{0} \right\rangle _p$ and $\left\vert \chi_{2}^{0} \right\rangle _n$ respectively describe the initial states of the
pointer (or other) particles inside the first and second measurement apparatuses. For
simplicity, we will treat these particules as unidimensional.

First assume that each source E$_{1,2}$ emits one photon; one is scattered, and the other is absorbed by either D$_{1}$ or D$_2$, depending on the state of the spin. An amplification process takes place in the photomultipliers, and drives the  positions of the $N$ particles inside the pointers of both apparatuses. After the
measurement has taken place, the state of the system becomes\footnote{We consider that, in the final state $\left\vert \Phi(t)\right\rangle$,  the two photons have been absorbed, either by one of the detectors, or the environment; one should then consider that the states $\left\vert \chi_{1,2} (t)\right\rangle $ also describes particles in the environment. The initial and final states of the radiation in (\ref{1}) and (\ref{2}) are then the vacuum state, which does not have to be written explicitly.
 Moreover, in order the increase the recoil effect of the atom, we may assume that each source has emitted $N$ photons instead of one. This would not change the structure of the calculation either.}:%
\begin{align}
\left\vert \Phi(t)\right\rangle & =\alpha\left\vert +\right\rangle
\left\vert \varphi_{z}^{+}(t)\right\rangle \left\vert
\varphi^0_{x}(t)\right\rangle \left\vert \varphi^0_{y}(t)\right\rangle
\prod \limits_{p=1}^{N}\left\vert \chi_{1}^{+}(t)\right\rangle_p \prod
\limits_{n=1}^{N}\left\vert \chi_{2}^{0}(t)\right\rangle_n  \notag \\
& \hspace{1cm} +\beta\left\vert -\right\rangle \left\vert \varphi_{z}^{-}(t)\right\rangle
\left\vert \varphi^0_{x}(t)\right\rangle \left\vert
\varphi^0_{y}(t)\right\rangle \prod \limits_{p=1}^{N}\left\vert
\chi_{1}^{0}(t)\right\rangle_p \prod \limits_{n=1}^{N}\left\vert
\chi_{2}^{-}(t)\right\rangle_n   \label{2}
\end{align}
In this expression, $\left\vert \varphi_{z}^{\pm}(t)\right\rangle $ is the
ket describing the motion of the atom wave packet of the particle along Oz
when a photon has provided a positive, or negative, recoil to the atom. In
both cases, we assume that the motion along $Ox$ and $Oy$ is not affected, and remains described by the freely propagating state $\left\vert \varphi^0_{x,y}(t)\right\rangle$.\
The wave functions associated with these kets are:%
\begin{equation}
\varphi_{z}^{\pm}(z,t)\sim\left[ a^{4}+\frac{4\hslash^{2}t^{2}}{m^{2}}\right]
^{-1/4}\exp\left\{ \pm i\frac{mv z}{\hslash}-\frac{\left[ z\mp v t%
\right] ^{2}}{a^{2}+\frac{2i\hslash t}{m}}\right\}   \label{3}
\end{equation}
and:%
\begin{equation}
\varphi_{x}^{0}(x,t)\sim\left[ a^{4}+\frac{4\hslash^{2}t^{2}}{m^{2}}\right]
^{-1/4}\exp\left\{ -\frac{x^{2}}{a^{2}+\frac{2i\hslash t}{m}}\right\}
\label{4}
\end{equation}
where $a$ is the minimal width of the Gaussian wave packet (assumed to be
the same along the three axes), $m$ the mass of the particle, and $\pm v$
its recoil velocity after absorbing a photon from either emitter E$_1$ or E$_2$. A similar expression gives the expression of the wave function along axis $Oy$.\
Since neither direction $Ox$ nor $Oy$ plays a role in the calculation, we merely ignore the corresponding wave functions in
what follows.\ The phase $%
S_{z}^{\pm}(z,t)$ of $\varphi_{z}^{\pm}(z,t)$ is:%
\begin{equation}
S_{z}^{\pm}(z,t)=\pm\frac{mv z}{\hslash}+\frac{2\hslash t}{m}\frac{\left[
z\mp v t\right] ^{2}}{a^{4}+\frac{4\hbar^{2}t^{2}}{m^{2}}}   \label{5}
\end{equation}

Similarly, the wave functions associated with the states $\vert \chi_{1,2}
^{0}(t)\rangle $ and $\vert
\chi_{1,2}^{+}(t)\rangle $ are\footnote{For the sake of simplicity, we assume that the arbitrary axis along which all pointer particles move is Oz; the variable $z_{p,n}$ gives the position of the $n,p$-th particle with respect to some initial reference position, which may differ from one particle to the other.}:%
\begin{align}
\chi^{0}(z,t) & \sim\left[ b^{4}+\frac{4\hslash^{2}t^{2}}{M^{2}}%
\right] ^{-1/4}\exp\left\{ -\frac{z^{2}}{b^{2}+\frac{2i\hslash t}{M}}%
\right\}  \notag \\
\chi^{\pm}(z,t) & \sim\left[ b^{4}+\frac{4\hslash^{2}t^{2}}{M^{2}}%
\right] ^{-1/4}\exp\left\{  \pm i\frac{MVz}{\hslash}-\frac{\left[ z \mp Vt%
\right] ^{2}}{b^{2}+\frac{2i\hslash t}{M}}\right\}   \label{6}
\end{align}
where M is the mass of the pointer particles, $b$ the initial width of
their wave packets, and $V$ their velocity when the detector has counted one
photon. Therefore, the phases $\xi^{0}(z,t)$ and $\xi^{+}(z,t)$ of these functions are:%
\begin{align}
\xi^{0}(z,t) & =\frac{2\hslash t z^{2}}{M\left[ b^{4}+4\frac{%
\hbar^{2}t^{2}}{M^{2}}\right] }  \notag \\
\xi^{\pm}(z,t) & = \pm \frac{MVz}{\hslash}+\frac{2\hslash t}{M}\frac{%
\left[ z \mp Vt\right] ^{2}}{\left[ b^{4}+4\frac{\hbar^{2}t^{2}}{M^{2}}%
\right] }   \label{6-2}
\end{align}

\section{Motion of the Bohmian positions}

We now calculate the motion of the Bohmian positions, driven by the
gradients of the phase of the wave functions.

\subsection{Equation of motion for a generic single particle with spin}

To obtain a first idea of the calculation, assume first that we have a single particle with spin $1/2$. It may be described at any time by the spinor:%
\begin{equation}
\alpha\left\vert +\right\rangle \left\vert
\varphi^{+}(t)\right\rangle +\beta\left\vert -\right\rangle \left\vert
\varphi^{-}(t)\right\rangle   \label{7}
\end{equation}
with:%
\begin{align}
\left\langle \mathbf{r}\right. \left\vert \varphi^{+}(t) \right\rangle &
=\varphi^{+}(\mathbf{r},t)= |\varphi^{+}(\mathbf{r},t)| \, \text{e}^{iS_{+}(\mathbf{r},t)}
\notag \\
\left\langle \mathbf{r}\right. \left\vert \varphi^{-}(t)\right\rangle &
=\varphi^{-}(\mathbf{r},t)=|\varphi^{-}(\mathbf{r},t)| \,
\text{e}^{iS_{-}(\mathbf{r},t)}
\label{8}
\end{align}
and $\left\vert \alpha\right\vert ^{2}+\left\vert \beta\right\vert ^{2}=1$; the two functions $\varphi^{\pm}(\mathbf{r},t)$ are supposed to be normalized.
The associated probability current is the sum of the contributions of the
two spin components:%
\begin{equation}
\mathbf{J}(\mathbf{r})=\frac{\hbar}{m}\left[ \left\vert \alpha\right\vert
^{2}\left\vert \varphi^{+}(\mathbf{r})\right\vert ^{2}\bm\nabla S_{+}(%
\mathbf{r})+\left\vert \beta\right\vert ^{2}\left\vert \varphi^{-}(%
\mathbf{r})\right\vert ^{2}\bm\nabla S_{-}(\mathbf{r})\right]   \label{9}
\end{equation}
The Bohmian velocity $\mathbf{v}(\mathbf{r})$ is nothing but the ratio
between this current and the local probability density:%
\begin{equation}
\mathbf{v}(\mathbf{r})=
\frac{\hbar / m}{\left\vert \alpha\right\vert ^{2}\left\vert \varphi^{+}(\mathbf{r})
\right\vert ^{2}+\left\vert \beta\right\vert ^{2}\left\vert \varphi^{-}(%
\mathbf{r})\right\vert ^{2}}
\left[ \left\vert \alpha
\right\vert ^{2}\left\vert \varphi^{+}(\mathbf{r})\right\vert ^{2}\bm\nabla S_{+}(\mathbf{r})
+ \left\vert
\beta\right\vert ^{2}\left\vert \varphi^{-}(\mathbf{r})\right\vert ^{2} \bm\nabla S_{-}(\mathbf{r})\right]   \label{10}
\end{equation}

\subsection{Coupling between the positions of the spin particle and pointer
particles}

We now come back to the optical Stern-Gerlach experiment.
From now on, and as mentioned above, we focus our calculation on the Oz component of the positions only.
 We call $Q$ the Oz component of the Bohmian position of the spin particle,
$Z_{n}$ and $Z_{p}$ the (one dimensional) Bohmian positions of the pointer particles.\ For the
many particle state (\ref{2}), the Bohmian velocity of the spin particle is:%
\begin{align}
\frac{\text{d}}{\text{d}t} Q & =\frac{\hbar}{mD(Q,Z_{p},Z_{n})}\left[
\left\vert \alpha\right\vert ^{2}\left\vert \varphi_{z}^{+}(Q,t)\right\vert
^{2}\prod \limits_{p}\left\vert \chi^{+}(Z_{p},t)\right\vert ^{2}\prod
\limits_{n}\left\vert \chi^{0}(Z_{n},t)\right\vert ^{2}\nabla
S_{+}(Q,t)\right.  \notag \\
& \hspace{3.5cm}\left. +\left\vert \beta\right\vert ^{2}\left\vert
\varphi_{z}^{-}(Q,t)\right\vert ^{2}\prod \limits_{p}\left\vert
\chi^{0}(Z_{p},t)\right\vert ^{2}\prod \limits_{n}\left\vert
\chi^{-}(Z_{n},t)\right\vert ^{2}\nabla S_{-}(Q,t)\right]   \label{11}
\end{align}
with:%
\begin{align}
D(Q,Z_{p},Z_{n})=\left\vert \alpha\right\vert ^{2}\left\vert
\varphi_{z}^{+}(Q,t)\right\vert ^{2} & \prod \limits_{p}\left\vert
\chi^{+}(Z_{p},t)\right\vert ^{2}\prod \limits_{n}\left\vert
\chi^{0}(Z_{n},t)\right\vert ^{2} \nonumber \\
& +\left\vert \beta\right\vert
^{2}\left\vert \varphi_{z}^{-}(Q,t)\right\vert ^{2}\prod
\limits_{p}\left\vert \chi^{0}(Z_{p},t)\right\vert ^{2}\prod
\limits_{n} \left\vert \chi_{2}^{-}(Z_{n},t)\right\vert ^{2}   \label{12}
\end{align}
and:%
\begin{equation}
\nabla S_{\pm}(Q,t)=\pm\frac{mv}{\hslash}+\frac{4\hslash t}{m}\frac{%
Q\mp v t}{\left[ a^{4}+\frac{4\hbar^{2}t^{2}}{m^{2}}\right] }
 \label{13}
\end{equation}
The velocity of the spin particle is therefore a weighted average of the
velocities associated with two wave packets at point $Q$, describing upwards and downwards motions resulting from opposite recoil effects. As expected with an entangled quantum state,
the weights depend, not only of the position $Q$ of the particle, but also
on the positions $Z_{p}$ and $Z_{n}$ of all pointer particles.

Similarly, the velocity of any pointer particle is given by the relation:%
\begin{align}
\frac{\text{d}}{\text{d}t}Z_{p} & =\frac{\hbar}{MD(Q,Z_{p},Z_{n})}\left[
\left\vert \alpha\right\vert ^{2}\left\vert \varphi_{z}^{+}(Q,t)\right\vert
^{2}\prod \limits_{p}\left\vert \chi^{+}(Z_{p},t)\right\vert ^{2}\prod
\limits_{n}\left\vert \chi^{0}(Z_{n},t)\right\vert
^{2}\nabla\xi^{+}(Z_{p},t)\right.
\notag \\
& \hspace{3.5cm}\left. +\left\vert \beta\right\vert ^{2}\left\vert
\varphi_{z}^{-}(Q,t)\right\vert ^{2}\prod \limits_{p}\left\vert
\chi^{0}(Z_{p},t)\right\vert ^{2}\prod \limits_{n}\left\vert
\chi^{-}(Z_{n},t)\right\vert ^{2}\nabla\xi^{0}(Z_{p},t)\right]
\label{14}
\end{align}
with:%
\begin{align}
\nabla\xi^{\pm}(Z,t) & = \pm \frac{MV}{\hslash}+\frac{4\hslash t}{M}\frac{%
Z \mp Vt}{\left[ b^{4}+4\frac{\hbar^{2}t^{2}}{M^{2}}\right] }  \notag \\
\nabla\xi^{0}(Z,t) & =\frac{4\hslash t Z}{M\left[ b^{4}+4\frac{%
\hslash^{2}t^{2}}{M^{2}}\right] }   \label{15}
\end{align}
where $Z$ stands for $Z_p$ or $Z_n$.
The time evolution of the positions $Z_n$ is given by an expression similar to (\ref{14}), where $\nabla\xi^{+}(Z_{p},t)$ is replaced by $\nabla\xi^{0}(Z_{n},t)$ and $\nabla\xi^{0}(Z_{p},t)$ by $\nabla\xi^{-}(Z_{n},t)$.
Again, we see
that the velocity of each pointer particle is a weighted average between the
velocities associated with two wave packets, but for the pointers one of the
wave packets is static.

\subsection{Slow and fast pointers}

The two  wave packets of the spin particle separate in a time of the order of $\tau_{a}\simeq a/ v_{z}$, those of the pointer particles separate in a time of the order of $\tau_{b}\simeq b/V$. The ratio between these times is:%
\begin{equation}
E=\frac{\tau_{a}}{\tau_{b}} = \frac{aV}{ b v}  \label{25}
\end{equation}
If $E>1$, the pointers are fast, indicating a result before the two wave packets of the spin particle separate; if $E<1$, they are slow, and the wave packets of the spin particle no longer overlap when the pointers start to indicate a definite result.

\subsection{The dynamics of all particles in each pointer reduces to that of a single particle}

We have assumed that all particles inside each of the two pointers have the same mass and are described by Gaussian wave packets with identical initial widths. In this case, we will show that their effect on the trajectory of the spin particle is identical to that of two fictitious pointers, each containing a single particle with  position $\hat{Z}_{1,2}$:
\begin{equation}\label{225}
\hat{Z}_{1}= \frac{1}{\sqrt N} \sum_{p=1}^N Z_{p}   \hspace{2cm}
\hat{Z}_{2}= \frac{1}{\sqrt N} \sum_{n=1}^N Z_{n}
\end{equation}
In these relations, the $\sqrt N$ ensures that $\hat{Z}_{1,2}$ have the same variance and statistical properties as any individual pointer position $Z_{n,p}$. In this substitution of variables, we will see that the effective velocity  associated with the fictitious  pointers becomes $\sqrt N V$.

To obtain these results, we insert relations (\ref{6})  into the expression (\ref{11}) of the velocity of the spin particle. We then see that several factors appear in both $\left\vert \chi_{2}^{0}(Z_{p,n},t)\right\vert ^{2}$ and $\left\vert \chi_{2}^{+}(Z_{p,n},t)\right\vert ^{2}$, cancelling each other in the numerator and the denominator. In (\ref{11}), we can therefore replace all the $|\chi^{0}(Z_{p,n},t)|^2$ by 1, and the $|\chi^{\pm}(Z_{p,n},t)|^2$ by the remaining factors:
\begin{equation}\label{substitutions-2}
\left\vert \chi^{\pm}(Z_{p,n},t) \right\vert ^2  \Rightarrow \exp  \left\{ \frac{ 2 b^2} {b^{4}+\frac{4\hslash ^2 t^2}{M^2}} \big[\pm 2 Vt Z_{p,n} - V^2 t^2 \big]\right\}
\end{equation}
Moreover, when the numbers of particles in both pointers are assumed to be equal, the terms in $-V^2 t^2$ appear $N$ times in both components in the numerator and the denominator, so that they also cancel each other. Taking for instance the product over $p$ then leads to the following substitution:
\begin{equation}\label{substitutions-2-bis}
\prod_p \left\vert \chi^{\pm}(Z_{p},t) \right\vert ^2  \Rightarrow
\exp \left\{ \pm \frac{4 b^2} {b^{4}+\frac{4\hslash ^2 t^2}{M^2}} \sqrt N   Vt  \hat{Z}_1\right\}
\end{equation}
A product over $n$ provides a similar result, where $\hat{Z}_1$ is replaced by $\hat{Z}_2$.

A similar simplification occurs with the wave function (\ref{3}) of the spin particle. The squared moduli of two wave functions of the spin particle are:
\begin{equation}\label{squared-modulus}
 \left\vert \varphi_{z}^{\pm}(Q,t) \right\vert ^2
 \sim \exp \left\{ - \frac{2 a^2}{a^4 + \frac{4 \hbar ^2 t^2}{m^2}}  \left[ Q \mp v t%
\right] ^{2} \right\}
\end{equation}
which also contain several common factors; only the terms linear in $vt$ are relevant. We can therefore make the substitutions:
\begin{equation}\label{substitutions-3}
 \left\vert \varphi_{z}^{\pm}(z,t) \right\vert ^2
 \Rightarrow
 \exp \left\{ \pm \frac{4 a^2  }{a^4 + \frac{4 \hbar ^2 t^2} {m^2} } v t Q \right\}
\end{equation}

We then obtain the simpler relations:
\begin{equation} \label{227}
\frac{\text{d}}{\text{d}t} Q  = \frac{\hbar}{m \overline{D}}\left[
\left\vert \alpha\right\vert ^{2}
\exp  \left\{ R_1  \right\}
\nabla
S_{+}(Q,t)
 +\left\vert \beta\right\vert ^{2}
\exp  \left\{ - R_2  \right\}
\nabla S_{-}(Q,t)
\right]
\end{equation}
with:
\begin{equation}\label{228}
\overline{D}=\left\vert \alpha\right\vert ^{2}
\exp  \left\{ R_1  \right\}
 +\left\vert \beta\right\vert
^{2}
\exp  \left\{ - R_2  \right\}
\end{equation}
and:
\begin{equation}\label{def-R}
R_{1,2}= \frac{4 a^2  }{a^4 + \frac{4 \hbar ^2 t^2} {m^2} } v t Q
+   \frac{4 b^2 } {b^{4}+\frac{4\hslash ^2 t^2}{M^2}}   \sqrt N   Vt \hat{Z}_{1,2}
\end{equation}
These equations show that the motion of $Q$ depends on the positions of the pointer particles only through the variables $\hat{Z}_1$ and $\hat{Z}_2$. One can actually even go further, and
note that the contribution of all pointer particles reduces to
that of the average position $(\hat{Z}_1 + \hat{Z}_2)/2$ of the fictitious pointers\footnote{The difference $(\hat{Z}_1 - \hat{Z}_2)$ evolves separately in time, under the only effect of the spreading of the wave packets.}.


Physically, for the first pointer, positive initial positions of the particles tend to favor the component of the state vector where the pointer wave packets have a positive velocity, over the other component in which they are static. If it turns out that the particles in the pointer already tend to indicate a positive result before the measurement starts, the probability of a positive spin result is increased. In the second pointer, positive initial positions  favor the same component of the state vector: in this component, the pointer wave packets have a zero initial velocity, while in the other they go further away with a negative velocity. So, for both pointers, positive values of the positions tend to favor trajectories of the spin particle flying upwards with a spin up. Moreover we have seen that, in the special case where all the wave packets are gaussian, only the sum of all values of $Z_n$ and $Z_p$ plays a role in the process.

In order to obtain the evolution of $\hat{Z}_1$, we now sum relation (\ref{14}) over all values of $p$ and divide by $\sqrt N $. As above, simplifications  between terms in the numerator and the denominator take place and we obtain:
\begin{equation} \label{229}
\frac{\text{d}}{\text{d}t} \hat{Z}_1   = \frac{\hbar}{M \overline{D}}\left[
\left\vert \alpha\right\vert ^{2}
\exp  \left\{ R_1  \right\}
\nabla \hat \xi^{+}(\hat{Z}_{1},t)
 +\left\vert \beta\right\vert ^{2}
\exp  \left\{ - R_2  \right\}
\nabla  \hat \xi^{0}(\hat{Z}_{1},t)
\right]
\end{equation}
with the values obtained from (\ref{15}):
\begin{align}
\nabla \hat \xi^{\pm}(\hat{Z},t) & =\pm \frac{M\sqrt N  V}{\hslash}+\frac{4\hslash t}{M}\frac{%
\hat{Z} \mp \sqrt N  Vt}{\left[ b^{4}+4\frac{\hbar^{2}t^{2}}{M^{2}}\right] }  \notag \\
\nabla \hat \xi^{0}(\hat{Z},t) & =\frac{4\hslash t \hat{Z}}{M\left[ b^{4}+4\frac{%
\hslash^{2}t^{2}}{M^{2}}\right] }   \label{155}
\end{align}
Similarly, we have:
\begin{equation} \label{229-2}
\frac{\text{d}}{\text{d}t} \hat{Z}_2   = \frac{\hbar}{M \overline{D}}\left[
\left\vert \alpha\right\vert ^{2}
\exp  \left\{ R_1  \right\}
\nabla \hat \xi^{0}(\hat{Z}_{2},t)
 +\left\vert \beta\right\vert ^{2}
\exp  \left\{ - R_2  \right\}
\nabla  \hat \xi^{-}(\hat{Z}_{2},t)
\right]
\end{equation}

As a result, because of the particular form of the Gaussian wave packets, all Bohmian positions inside each  pointer can be replaced by the position of their center of mass multiplied by $\sqrt N$, and be treated as a the position of a single fictitious pointer particle. The equations of motion (\ref{227}), (\ref{229}) and (\ref{229-2})  show that the velocity of the fictitious pointers is also multiplied by the square root $\sqrt N$.  If $N$ is very large, the fictitious pointers become very fast;  we will see that this implies that the influence of the pointers on the trajectory of the spin particle is dominant.


%
%
%
%


\subsection{Dimensionless variables}

For convenience, and use in the numerical calculations, we  introduce dimensionless variables by setting:
\begin{equation}
Q^{\prime}=\frac{Q}{a} \hspace{2.2cm}v^{\prime}=\frac {ma}{%
\hbar}v_{z}   \label{16}
\end{equation}
All positions are then expressed in terms of the initial widths of the
corresponding wave packets. For the pointer particles, we set:%
\begin{equation}
Z_{p,n}^{\prime}=\frac{Z_{p,n}}{b}   \hspace{2cm}
V^{\prime}=\frac{Mb}{\hbar}V    \label{20}
\end{equation}
and:%
\begin{equation}
   \label{21}
   \hat{Z}_{1,2}^{\prime} = \frac{\hat{Z}_{1,2}}{b}
\end{equation}
 We also introduce a dimensionless time $t^{\prime}$
 by:%
\begin{equation}
t^{\prime}=\frac{\hbar}{ma^{2}}t  \label{art-7}
\end{equation}
and note that the rapidity parameter $E$ is now given by:
\begin{equation}\label{rapidity}
  E=\eta \frac{V^{\prime}}{v^{\prime}}
\end{equation}
  with:
  \begin{equation}\label{17-3}
\eta =\frac{m a^{2}}{M b^{2}}
\end{equation}
In (\ref{227}) and following equations, we can then make the substitutions:
\begin{equation}\label{17-modif}
\exp \left\{ \pm \frac{4 a^2 }{a^4 + \frac{4 \hbar ^2 t^2} {m^2} }
v t Q \right\}
\Rightarrow
\exp \left\{ \pm \frac{4 v' t' }{1 + 4 (t')^2 } v' t' Q'
 \right\}
\end{equation}
 and:
 \begin{equation}\label{17-2}
  \exp  \left\{ \pm \frac{4 b^2} {b^{4}+\frac{4\hslash ^2 t^2}{M^2}} \sqrt N  Vt   \hat{Z}_{1,2}  \right\}
=
   \exp  \left\{ \pm \frac{4 \eta \sqrt N V' t' \hat{Z}_{1,2} ' } {1+ 4 \eta^2 (t')^2 }   \right\}
  \end{equation}
For the spin particle, the gradients of the phase are given by:%
\begin{equation}\label{19}
 \nabla S_{z}^{\pm}(Q)  =
 \pm\frac{v^{\prime}}{a}+\frac{4 t^{\prime}}{a}\frac{Q^{\prime}\mp v^{\prime
}t^{\prime}}{ 1+4(t^{\prime})^{2} }
= \frac{1}{a} ~ \frac{ \pm v' + 4 t' Q' }{1 + 4 (t')^2 }
\end{equation}
while, for the fictitious pointer particles, relations (\ref{155}) lead to:
\begin{align} \label{19-2}
\nabla\xi_{1}^{\pm}  & = \pm \frac{\sqrt N  V'}{b}+\frac{4 \eta t'}{b}\frac{%
\hat{Z}' \mp \eta \sqrt N V'  t'}{ 1+ 4 \eta^2 (t')^2 }   \notag \\
\nabla\xi_{1}^{0}  & = \frac{4 \eta t^{\prime}}{b} \frac{ \hat{Z}'}{
1+ 4 \eta^{2}(t^{\prime})^{2} }
\end{align}

When these substitutions are made, we obtain dimensionless  equations of evolution:
\begin{equation} \label{19-3}
\frac{\text{d}}{\text{d}t'} Q'  =\frac{1}{\overline{D}'}\left[ \left\vert \alpha \right\vert^2
\exp  \left\{ R_1 '
\right\}
\Big[\frac{v^{\prime
} + 4  t^{\prime}Q^{\prime}}{ 1+4(t^{\prime})^{2} } \Big]
 +  \left\vert \beta \right\vert^2
\exp  \left\{ - R_2 '
\right\}
\Big[\frac{- v^{\prime
} + 4  t^{\prime}Q^{\prime}}{ 1+4(t^{\prime})^{2} } \Big]
\right]
\end{equation}
with:
\begin{equation} \label{19-5}
\overline{D}'  =\left[ \left\vert \alpha \right\vert^2
\exp  \left\{ R_1 '
\right\}
 +  \left\vert \beta \right\vert^2
\exp  \left\{ - R_2 '
\right\}
\right]
\end{equation}
and:
\begin{equation}\label{def-R-prime}
R_{1,2}'=  \frac{4 v' t' Q' }{1 + 4 (t')^2 }
+  \frac{4\eta \sqrt N V' t' \hat{Z}_{1,2} ' } {1+ 4 \eta^2 (t')^2 }
\end{equation}
For the motion of the pointers, we obtain:
\begin{align} \label{19-4}
\frac{\text{d}}{\text{d}t'} \hat{Z}^{\prime}_1 & =\frac{1}{\overline{D}'}\left[ \left\vert \alpha \right\vert^2
\exp  \left\{ R_1 '
\right\}
\Big[ \frac{\sqrt N  V'+4\eta  t'
\hat{Z}_1 ' }{ 1+ 4 \eta (t')^2 }  \Big]
\right.
\notag \\
& \hspace{5cm}\left. +
 \left\vert \beta \right\vert^2 \exp  \left\{ - R_2 '
\right\}
\Big[  \frac{ 4 \eta t^{\prime}  \hat{Z}_1 '}{
1+ 4 \eta^{2}(t^{\prime})^{2} } \Big]
\right]
\end{align}
and:
\begin{align} \label{19-4-bis}
\frac{\text{d}}{\text{d}t'} \hat{Z}^{\prime}_2 & =\frac{1}{\overline{D}'}\left[ \left\vert \alpha \right\vert^2
\exp  \left\{ R_1 '
\right\}
\Big[ \frac{4 \eta t^{\prime} \hat{Z}_2 '}{
1+ 4 \eta^{2}(t^{\prime})^{2}} \Big]
\right.
\notag \\
& \hspace{1.5cm}\left. +
 \left\vert \beta \right\vert^2 \exp  \left\{ - R_2 '
\right\}
\Big[    \frac{  \sqrt N  V' + 4\eta t'
\hat{Z}_2 ' }{ 1+ 4 \eta^2 (t')^2 }  \Big]
\right]
\end{align}

We notice that the Planck constant $\hbar$ has disappeared from these equations. In addition to the initial degree of polarization of the spin:
\begin{equation}\label{dege-polar}
\sigma = \left\vert \alpha \right\vert ^2 - \left\vert \beta \right\vert ^2
\end{equation}
these equations contain four dimensionless parameters: the two reduced velocities $v'$ and $V'$, the parameter $\eta$, and the number of particles $N$ within each pointer. Since $N$ and $V'$ appear only through the product $\sqrt N V$, the number of these four parameters actually reduces to three.

\section{Results}
In order to understand the influence of the values of the initial positions of the spin particle and those of the pointers, we now apply the preceding considerations to various situations.

\subsection{Zero initial spin polarization}

 We first study the case where the initial spin polarization is zero, meaning that no result of measurement is privileged by the rules of standard quantum mechanics. Any imbalance can then result only from the influence of the Bohmian positions, either of the spin particle, or of the particles in the pointers.

\subsubsection{Uncoupled pointers, or no pointer}

\begin{figure}[t]   
\begin{minipage}[c]{0.32\textwidth}
\includegraphics[width=4cm]{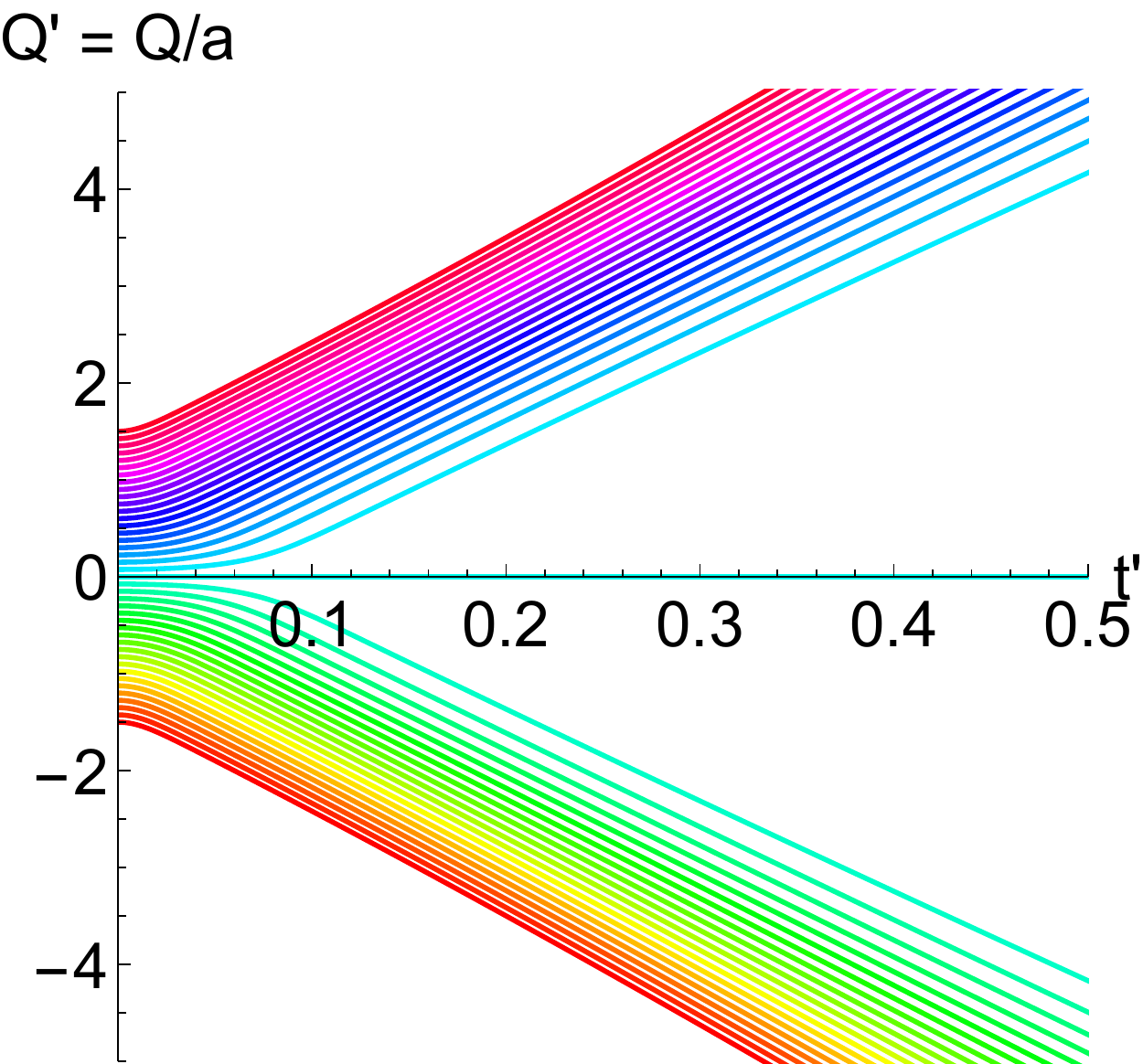}
\end{minipage}
\hfill
\begin{minipage}[c]{0.32\textwidth}
\includegraphics[width=4cm]{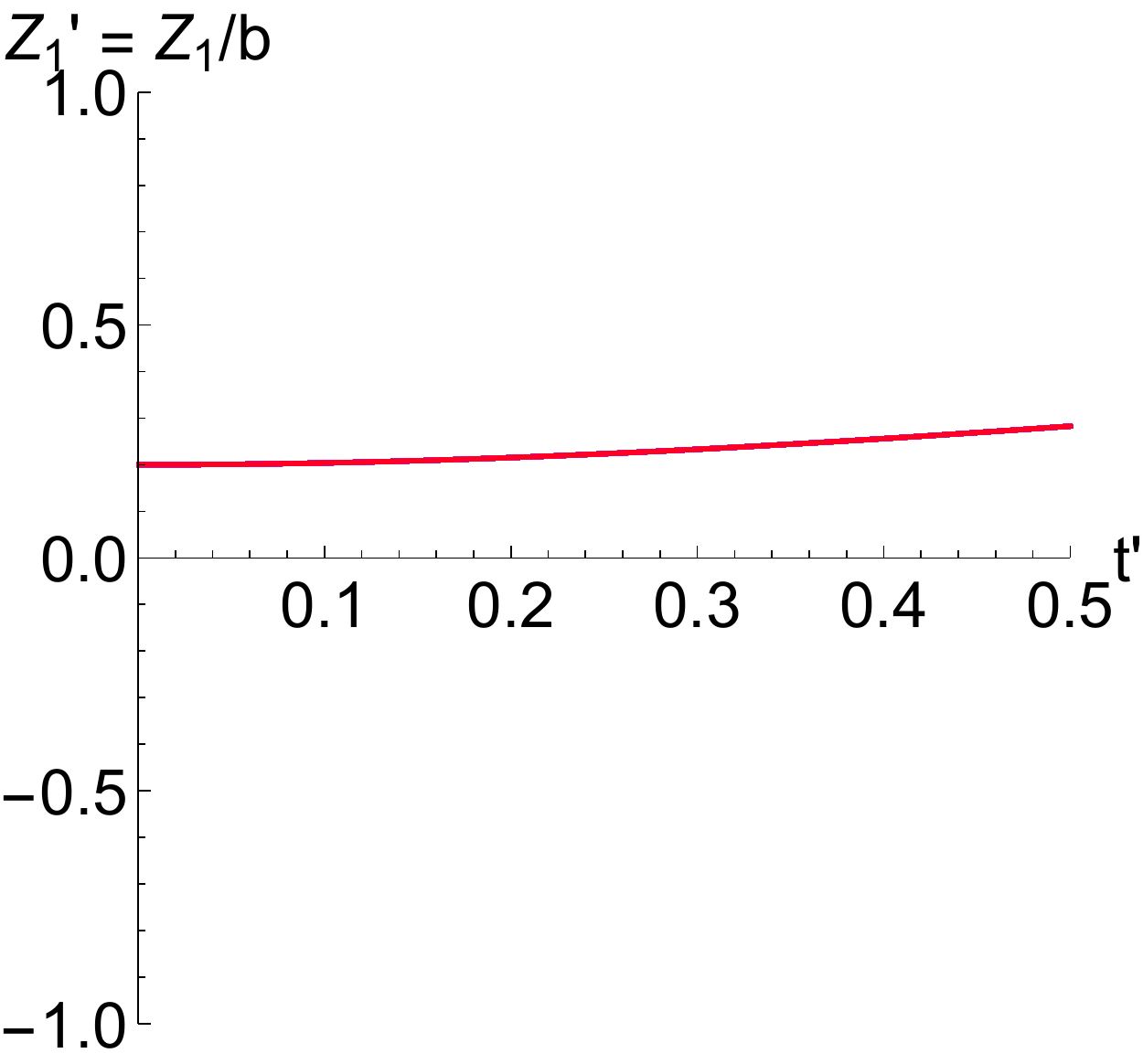}
\end{minipage}
\hfill
\begin{minipage}[c]{0.32\textwidth}
\includegraphics[width=4cm]{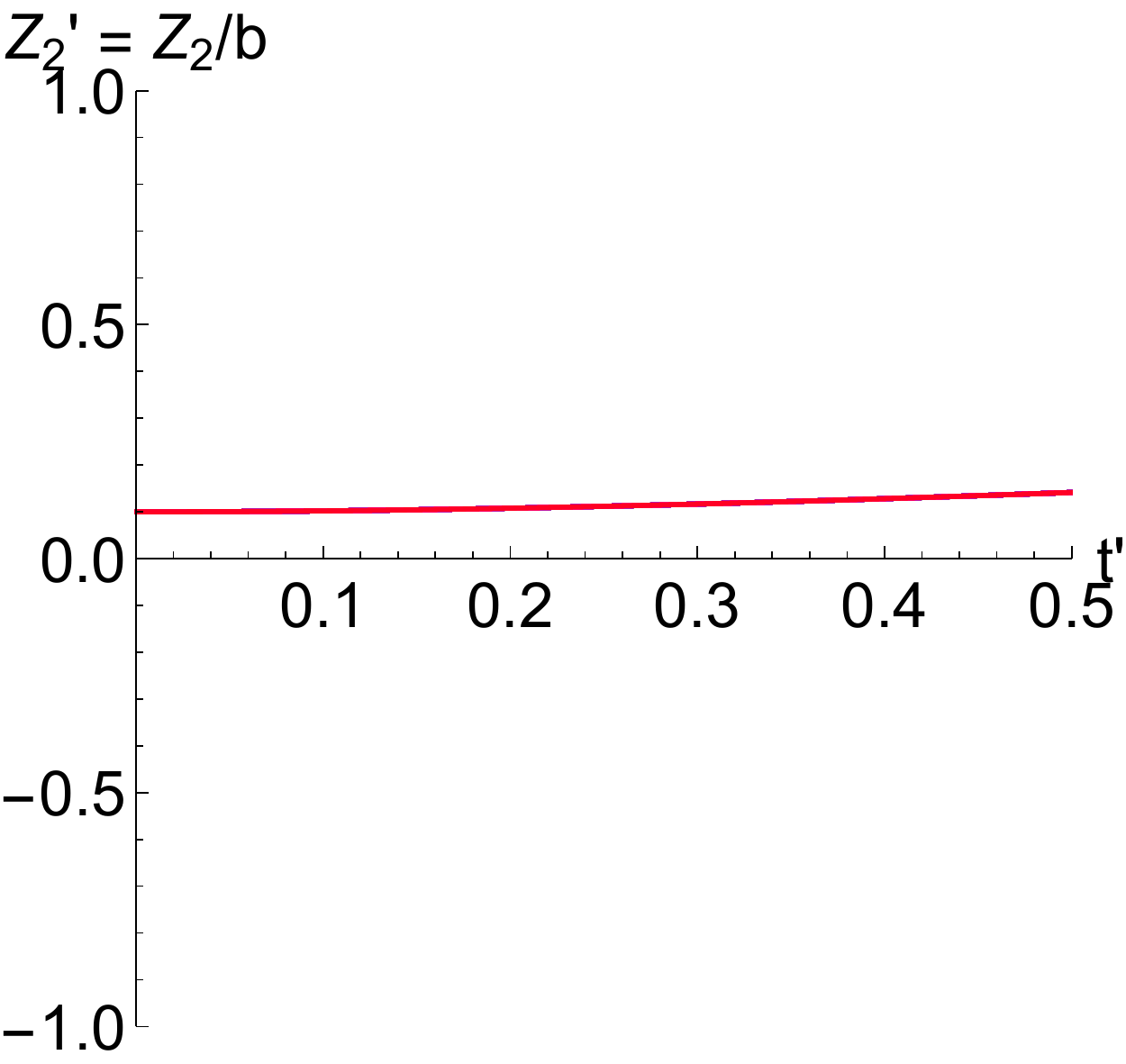}
\end{minipage}
 \bigskip
\caption{Trajectories obtained when $V' =0$, so that the motion of the pointers decouples
 from that of the spin particle. Each pointer contains only one particle. The initial spin polarization of the spin particle is $\sigma = 0$. In all figures of this article, we take the parameter $\eta=ma^2 /M b^2$ equal to unity ($\eta =1$) and $v'=ma v_z / \hbar = 10$. \newline The left part of the figure shows the trajectories of the spin particle starting from different initial positions, the central and right part the trajectories of the particles in the first and second pointers. Since $V' =0$, the trajectories of the spin particles are exactly the same as those usually obtained when the measurement apparatus is treated classically (through a given external  potential). The initial positions of the two pointer particles are $Z_1 (0)=0.2$ and $Z_2 (0) = 0.1$, but these values are irrelevant for the trajectory of the spin particle. The pointers remain motionless, except a small drift due to the spreading of their (otherwise static) wave packets.}
\label{fig-2}
\vspace{2mm}
\hrule
\end{figure}

If we set $V'=0$ in the equations, $R_1 '$ and $R_2 '$ are equal, and the motion of the spin particle decouples from that of the pointers. Physically, when  the pointer states $\chi^{0}(z,t)$ and $\chi^{\pm}(z,t)$ in (\ref{6}) remain identical, no
 quantum entanglement develops from the initial quantum state (\ref{1}). Figure \ref{fig-2} shows the trajectories obtained in this case. They are actually exactly the same as the usual trajectories (when the
 effect of the measurement apparatus is treated as a classical external potential). For simplicity, we have assumed that each pointer contains a single particle, but
  the value of $N$ is actually irrelevant for the motion of the spin particle, which remains unaffected by the pointers. The left part of the figure shows several
    trajectories of the spin particle starting from various initial positions (41 different values of $Q'(0)$, equally spaced in the interval $\pm 1.5$) , the central part the trajectory of the first pointer, and the right part the
    trajectory of the second pointer. As expected in the absence of entanglement, both pointers remain still in this case, and therefore do  not indicate any result of measurement.

\subsubsection{Few particles in the pointers}

\begin{figure}[t]  
\begin{minipage}[c]{0.32\textwidth}
\includegraphics[width=4cm]{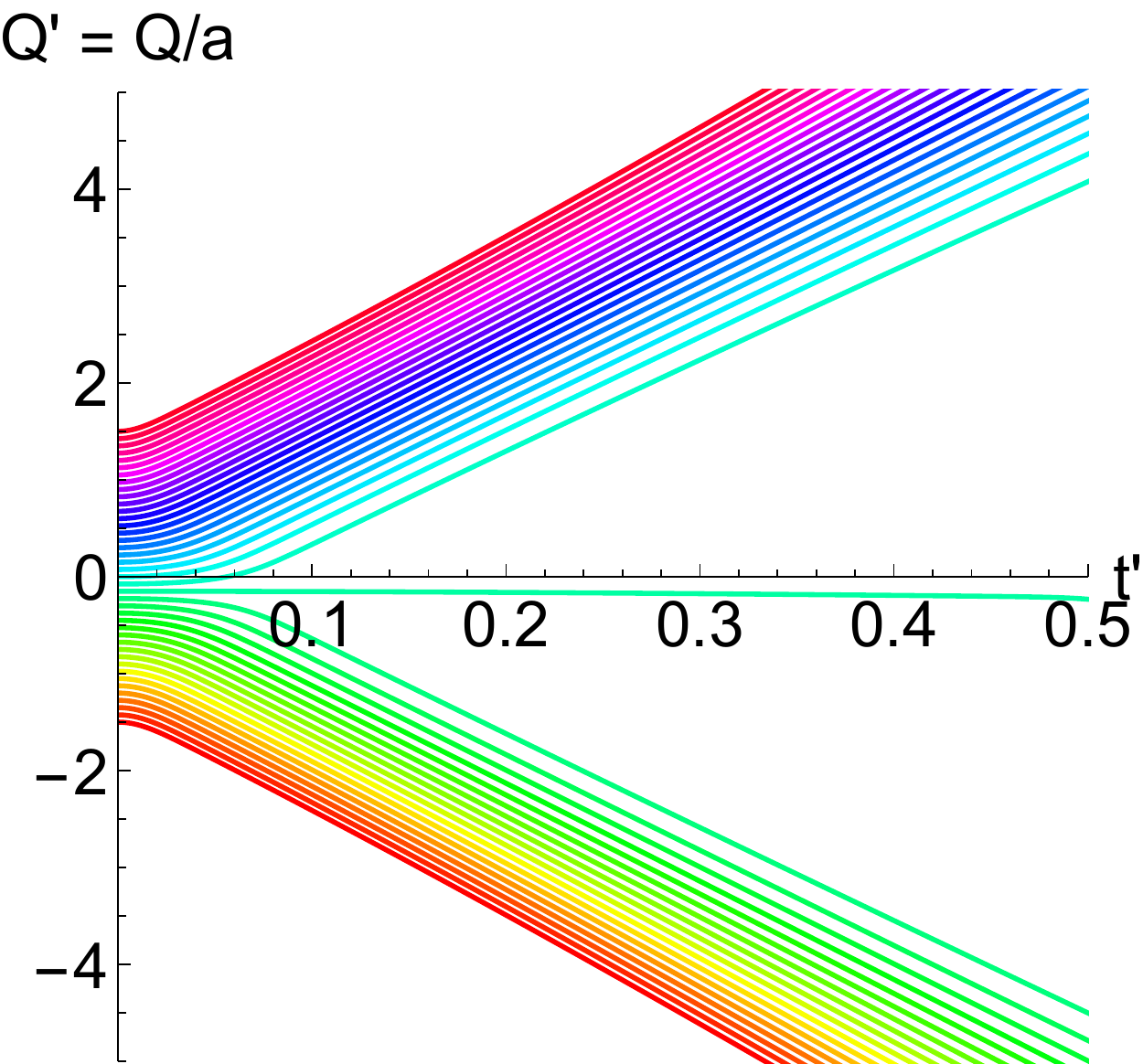}
\end{minipage}
\hfill
\begin{minipage}[c]{0.32\textwidth}
\includegraphics[width=4cm]{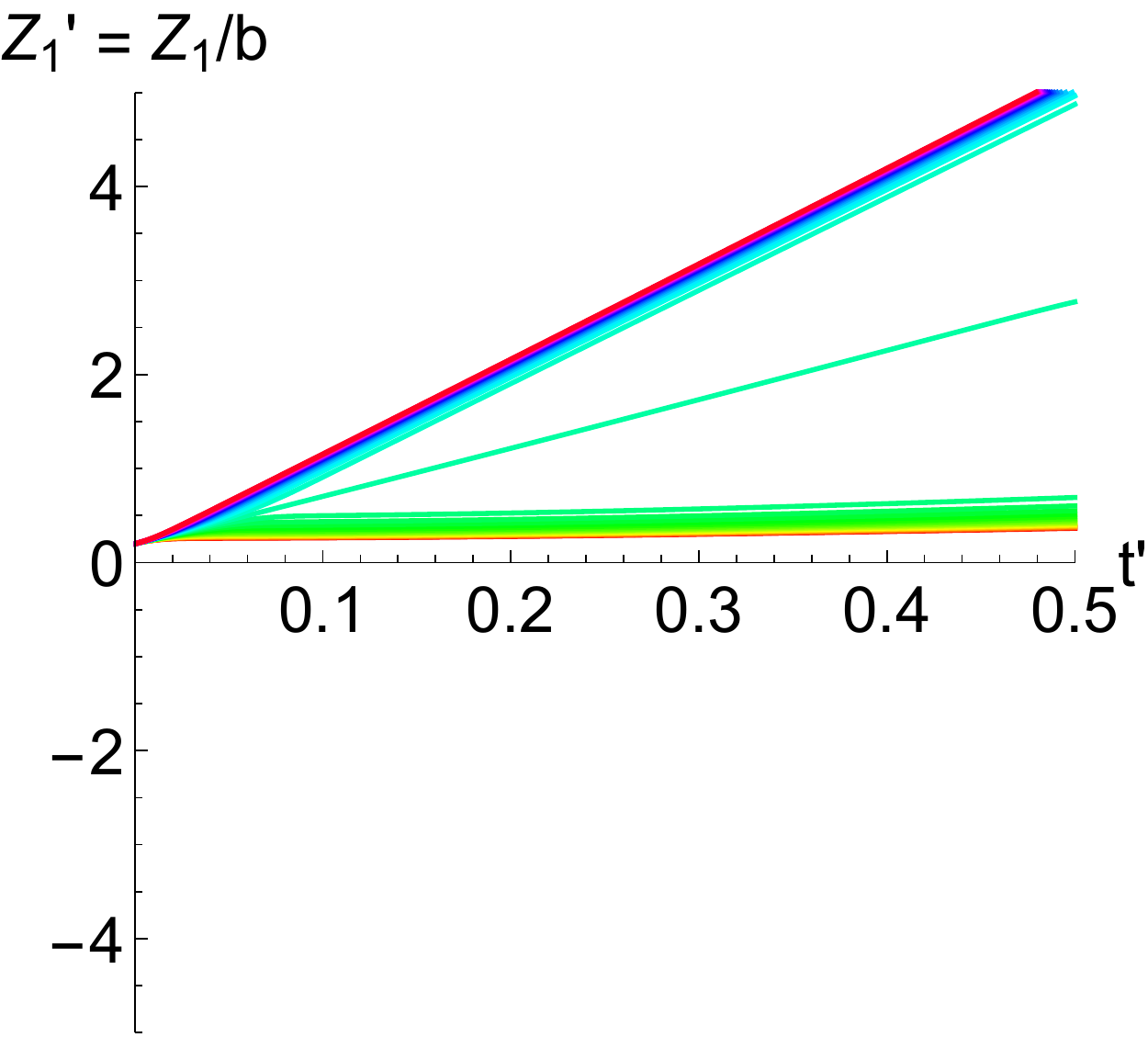}
\end{minipage}
\hfill
\begin{minipage}[c]{0.32\textwidth}
\includegraphics[width=4cm]{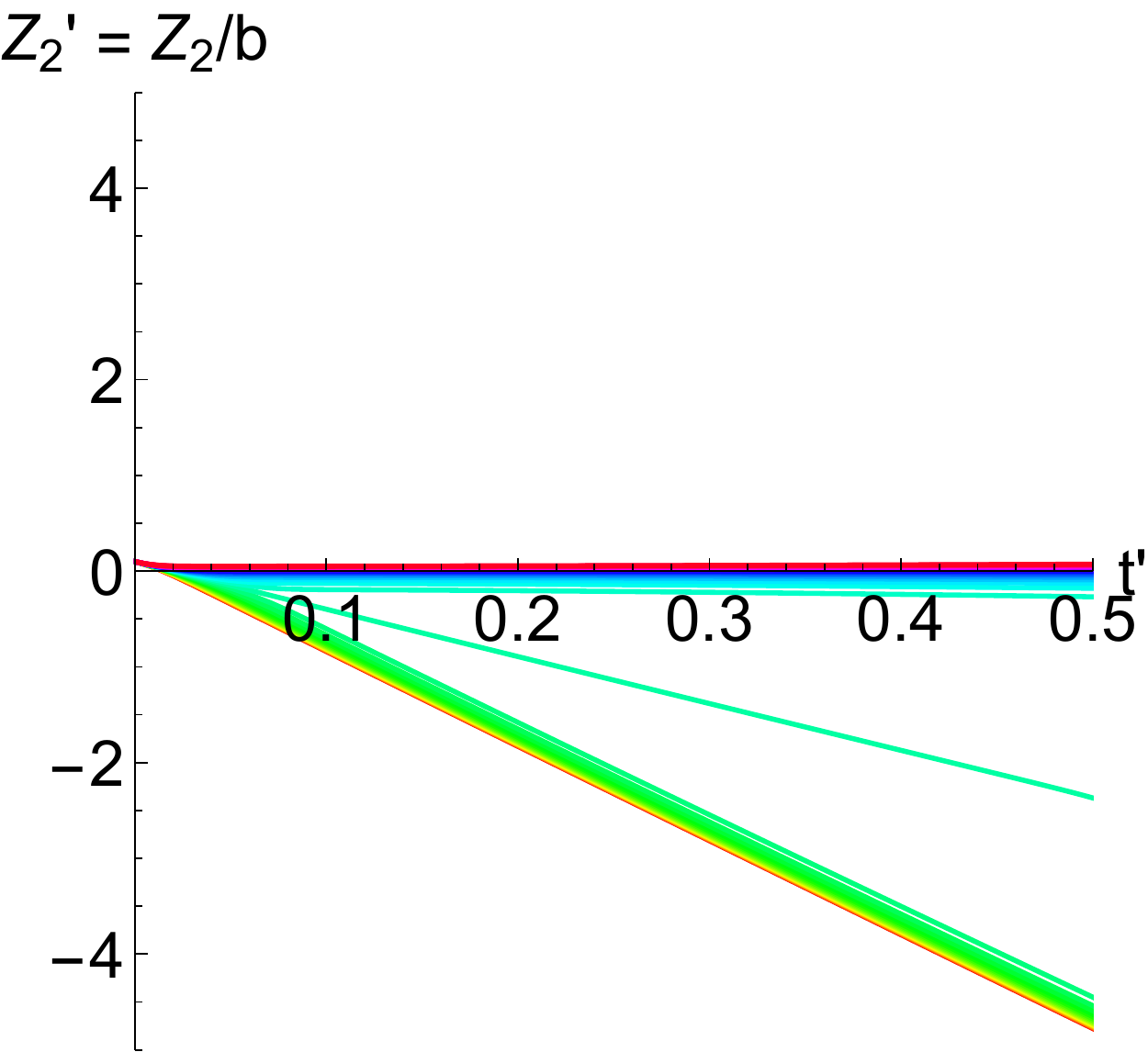}
\end{minipage}
 \bigskip
\caption{The only change of the input parameters with respect to those of figure~\ref{fig-2} is that $V'/v' = 1$, so that the motions of the spin particle and  of the pointers are  coupled. The positions of the two pointer particles now significantly evolve in time: if the spin particle flies upwards at the end of the measurement, the first pointer (central part of the figure) moves upwards, while the second pointer (right part of the figure) remains still; if the spin particle flies downwards, the first pointer remains still and the second pointer moves downwards. The trajectories of the spin particle remain similar to those in figure~\ref{fig-2}.
}
\label{fig-3}
\vspace{2mm}
\hrule
\end{figure}

Now assume that the pointers have a finite velocity ($V'=v'$), so that the pointers are entangled with the spin particle. Each pointer still contains only one particle, with the same initial position as in figure~\ref{fig-2}. The trajectories are shown in figure~\ref{fig-3}.
The major change is that the pointers now move and indicate in which direction the spin particle flies at the end of the measurement process. If the spin
particle flies upwards, the trajectory of the first pointer moves upwards, while that of the second pointer remains almost motionless; if the spin particle flies
downwards, the first pointer remains still while  and the second pointer moves downwards. Nevertheless, with only two pointer particles coupled to the spin particle, and with the initial positions  $Z_1 (0)=0.2$ and $Z_2 (0) = 0.1$, the trajectories of
the spin particle are not strongly changed with respect to those of figure~\ref{fig-2}.

\begin{figure}[!b]  
\hrule
\vspace{2mm}
\centering
\begin{minipage}[c]{0.32\textwidth}
\includegraphics[width=4cm]{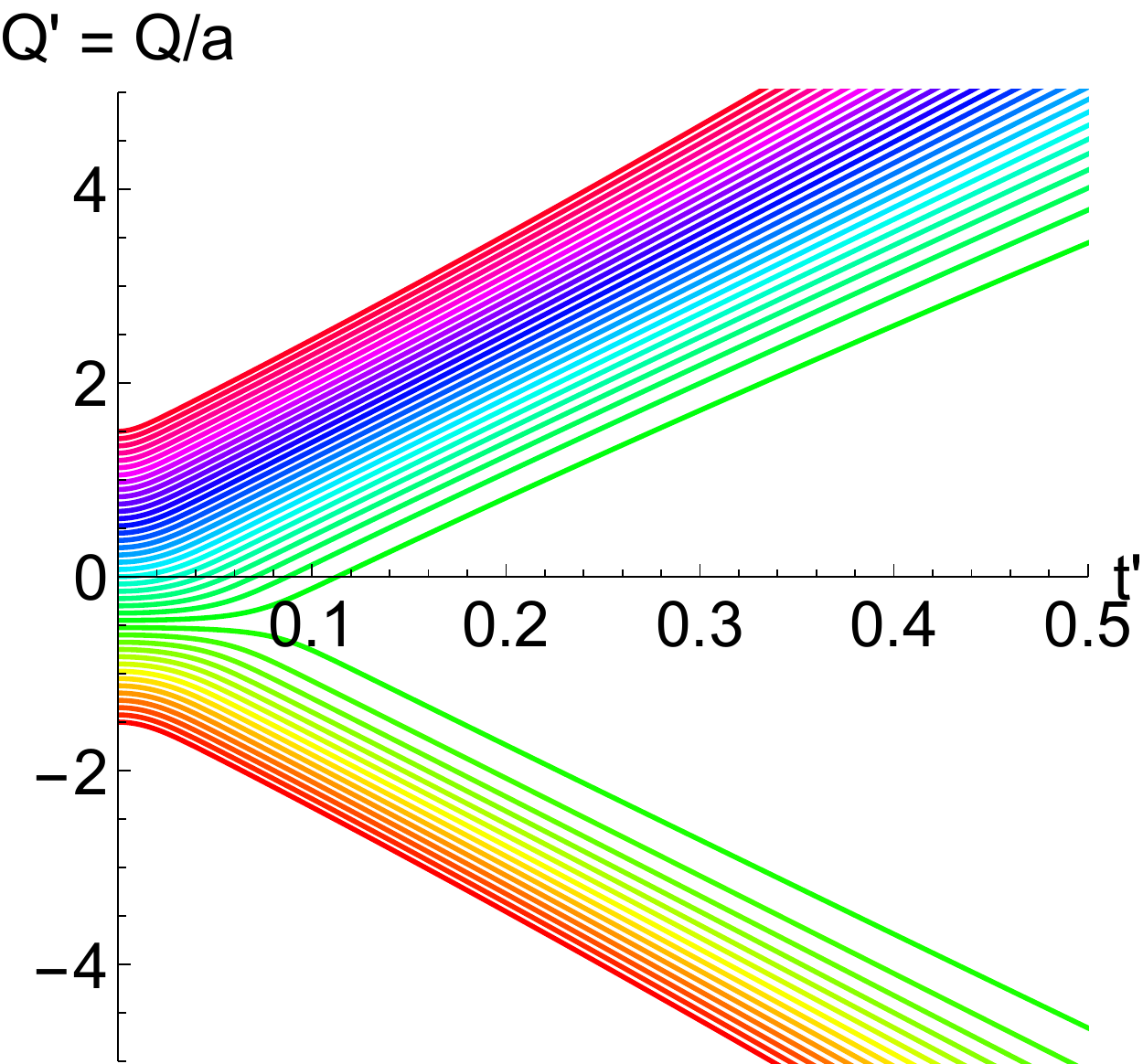}
\end{minipage}
\hfill
\begin{minipage}[c]{0.32\textwidth}
\includegraphics[width=4cm]{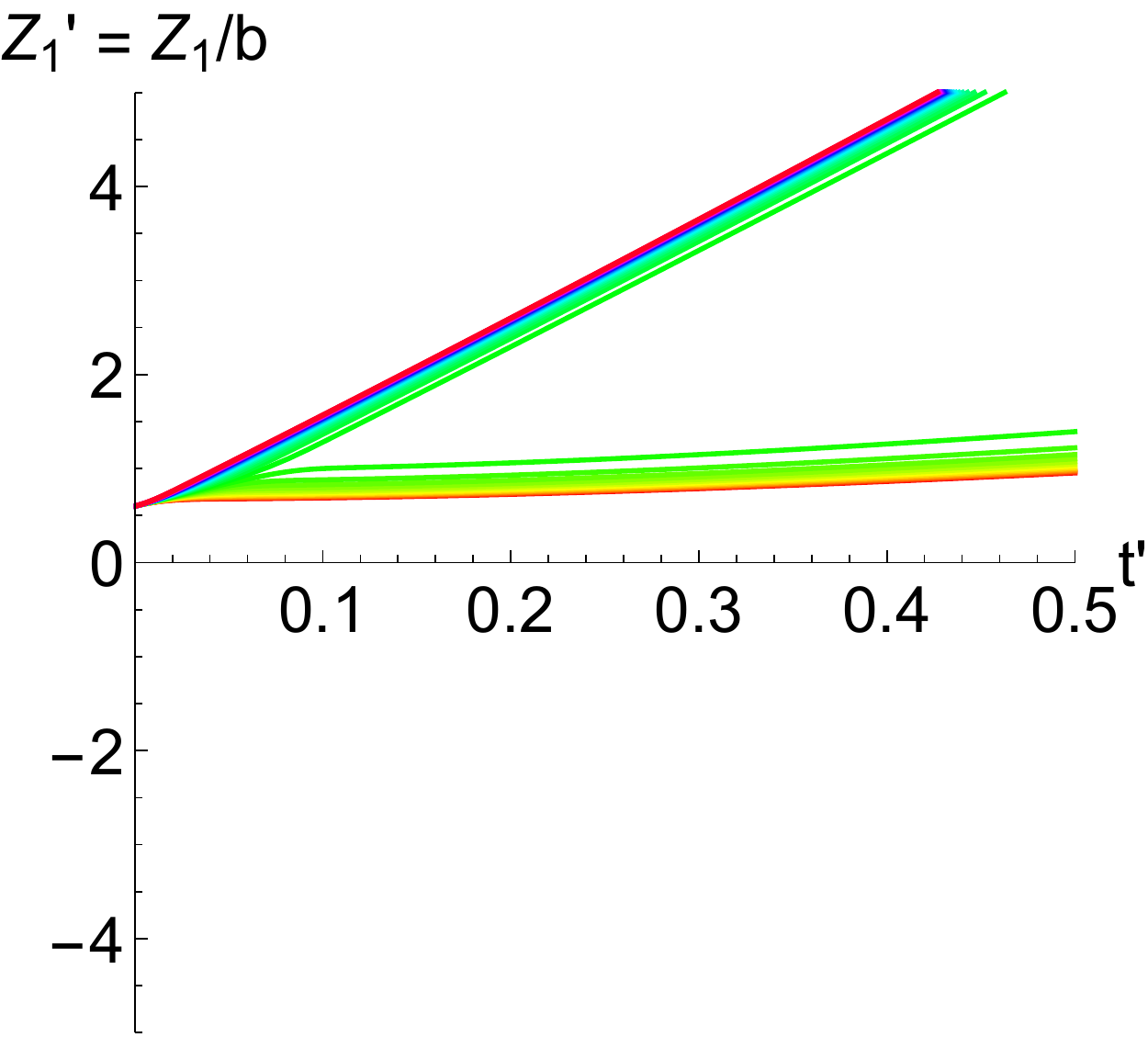}
\end{minipage}
\hfill
\begin{minipage}[c]{0.32\textwidth}
\includegraphics[width=4cm]{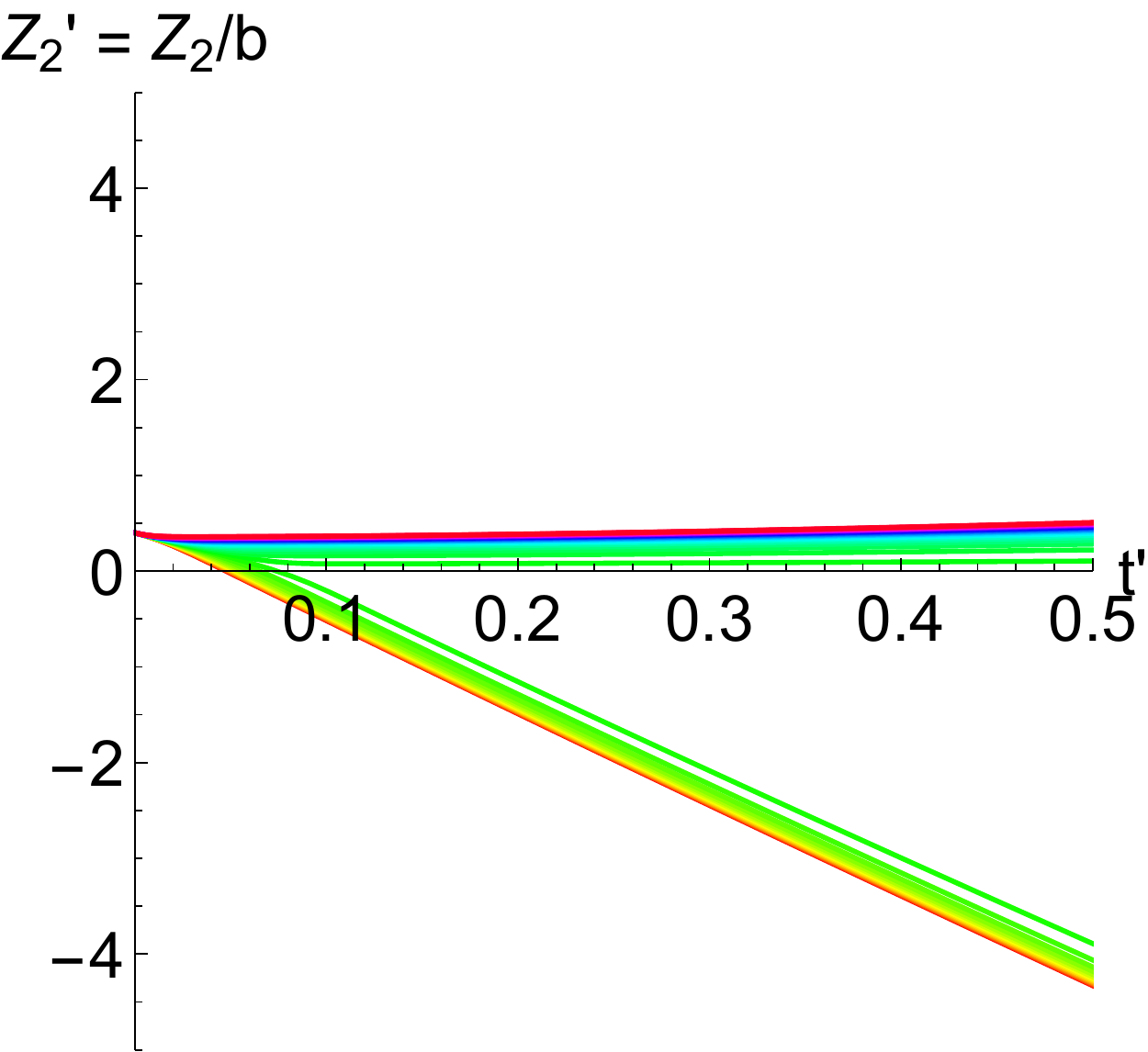}
\end{minipage}
 \bigskip
\caption{Same situation as in figure~\ref{fig-3} (no spin polarization, two pointer particles), but now the initial positions of the pointer particles have larger positive $Z_1 (0) = 0.6$ and $Z_2 (0) = 0.4$. This favors an upwards deviation of the spin  particle.  Even if the quantum state is exactly the same as in figure~\ref{fig-3}, the positive initial values of the positions of the two pointer particles have a significant effect on the trajectories of the spin particle.}
\label{fig-4}
\end{figure}

\begin{figure}[t]  
\centering
\begin{minipage}[c]{0.32\textwidth}
\includegraphics[width=4cm]{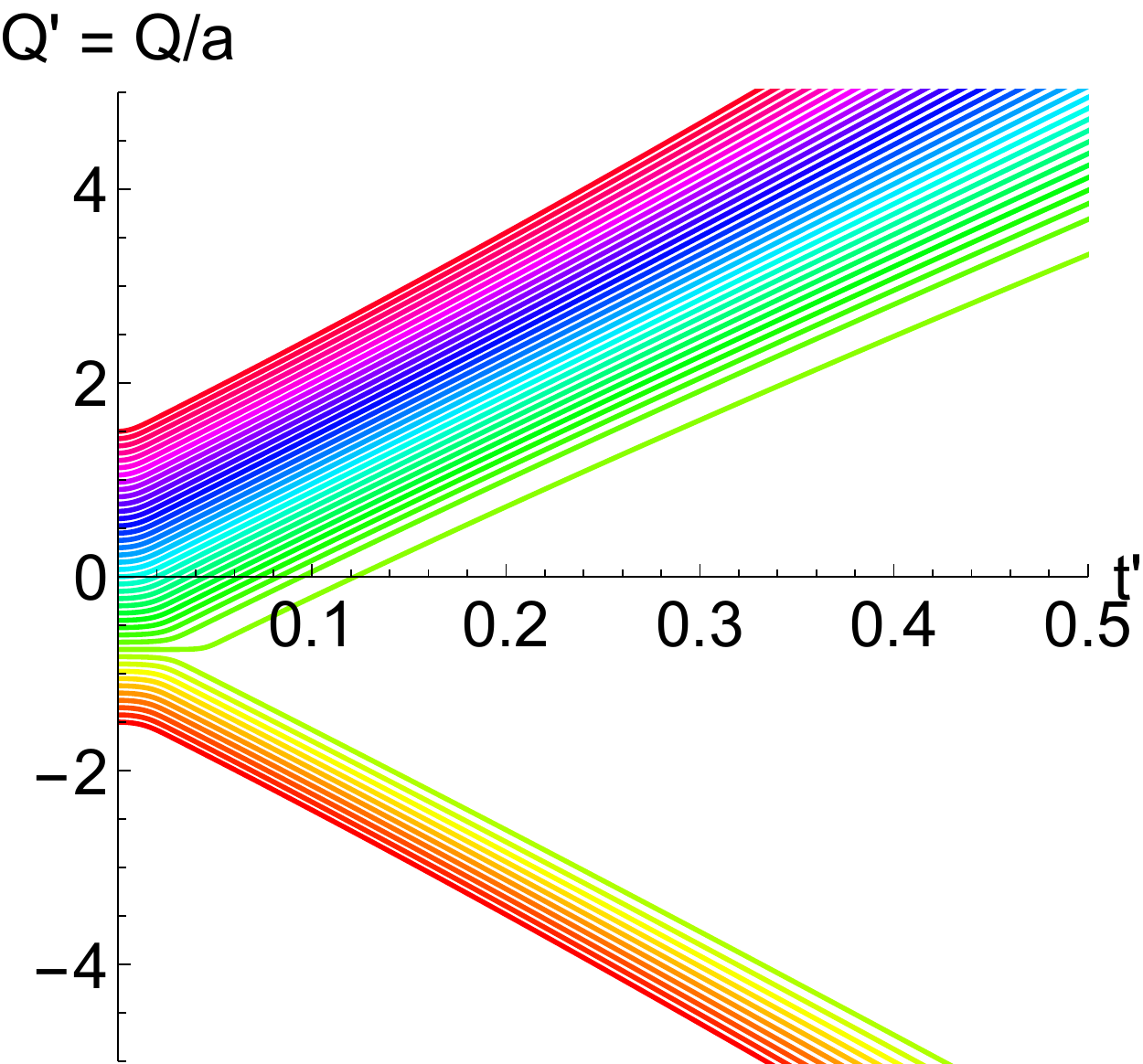}
\end{minipage}
\hfill
\begin{minipage}[c]{0.32\textwidth}
\includegraphics[width=4cm]{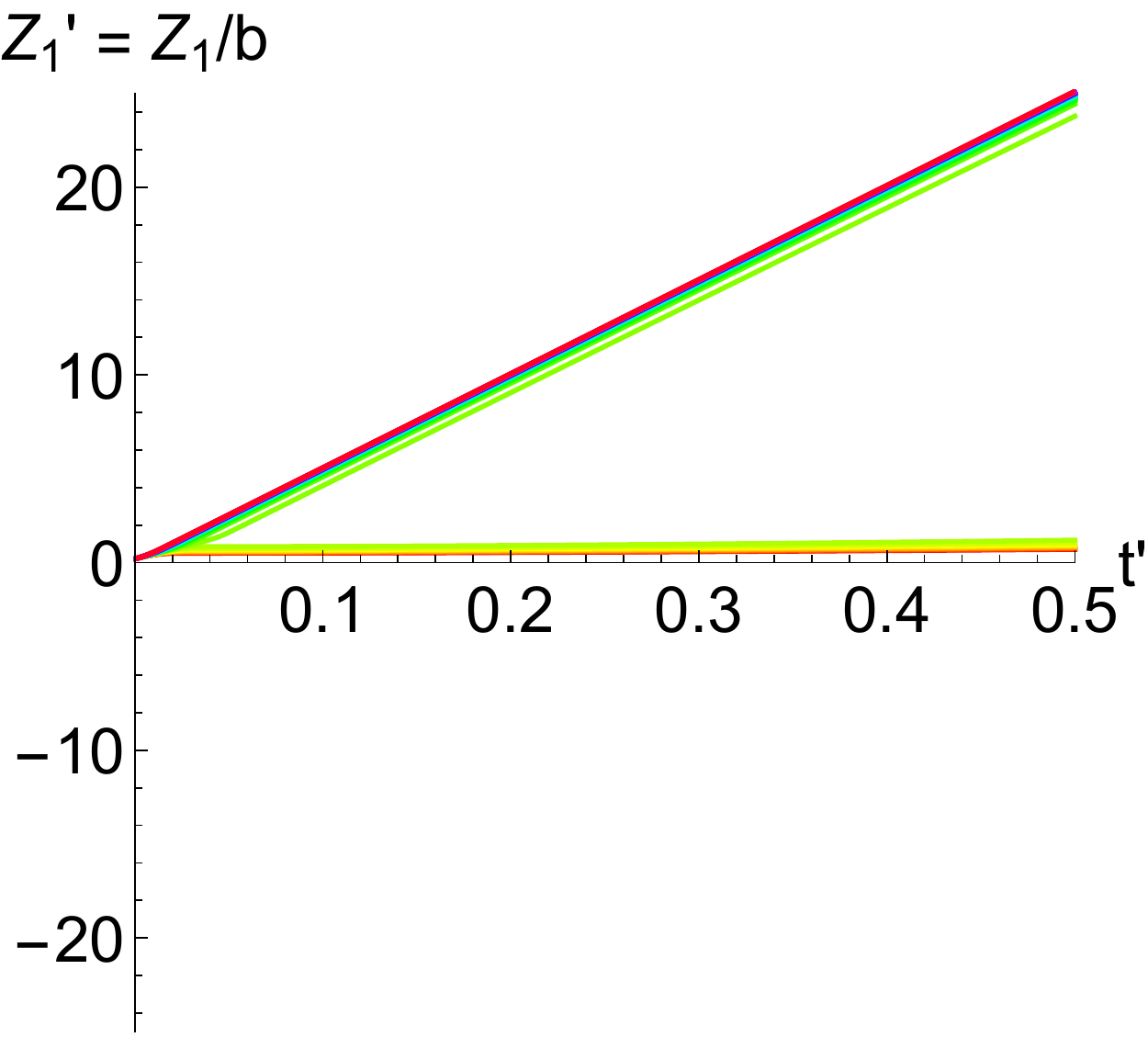}
\end{minipage}
\hfill
\begin{minipage}[c]{0.32\textwidth}
\includegraphics[width=4cm]{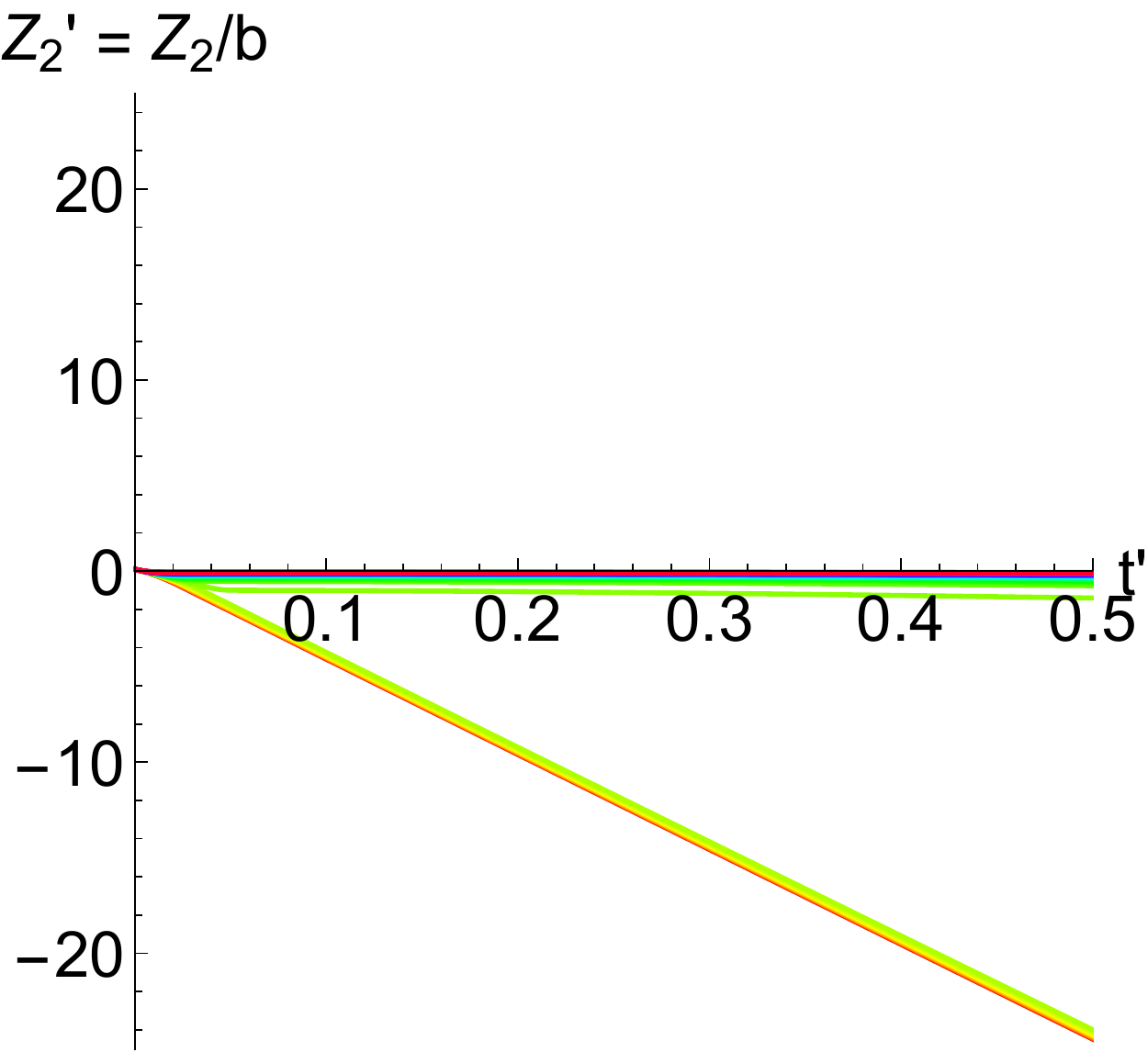}
\end{minipage}
 \bigskip
\caption{Same situation as in figure~\ref{fig-3} (no spin polarization, $\hat{Z}_1 (0)=0.2$ and $\hat{Z}_2 (0) = 0.1$), but with 25 particles in each pointer.
The  increase of the number of particles in the pointers significantly  enhances their influence on the trajectory of the spin particle.}.
\label{fig-5}
\hrule
\vspace{2mm}
\end{figure}

Now, if the initial positions of the pointers take larger positive values, which favours a spin up result, figure~\ref{fig-4} shows that more trajectories of the spin particle go in
 the up direction. This is a first illustration of the influence of the initial positions of the particles in the pointers: the proportion of trajectories going up and down differs from the simple predictions of the Born rule. One could then worry because, in this figure, the proportion of trajectories going up is larger than the quantum probability of obtaining a spin up result. Nevertheless, in other situations where the initial positions of
  the pointers favour a spin down result, the opposite is true (the trajectories are merely obtained from those of  figure~\ref{fig-4} by a symmetry with respect
  to the horizontal axis). If a proper quantum average is taken over the initial positions of the pointers, the proportion of trajectories going up and down reproduces the quantum average exactly.
Not surprisingly, the effect of the initial positions of the pointers increases with larger values of the number of pointer particles. Figure~\ref{fig-5} shows what happens when $N$ increases from $1$ to $25$. We see that this significantly enhances the influence of the pointer particles on the Bohmian trajectory of the spin particle (and spin direction at the end of the measurement).

\subsubsection{Pointers containing many particles}
\label{many-pointers-1}

In a realistic measurement apparatus, the pointers are macroscopic objects, and the number of particles they contain is some fraction of the Avogadro number.

\begin{figure}[t] 
\centering
\begin{minipage}[c]{0.32\textwidth}
\includegraphics[width=4cm]{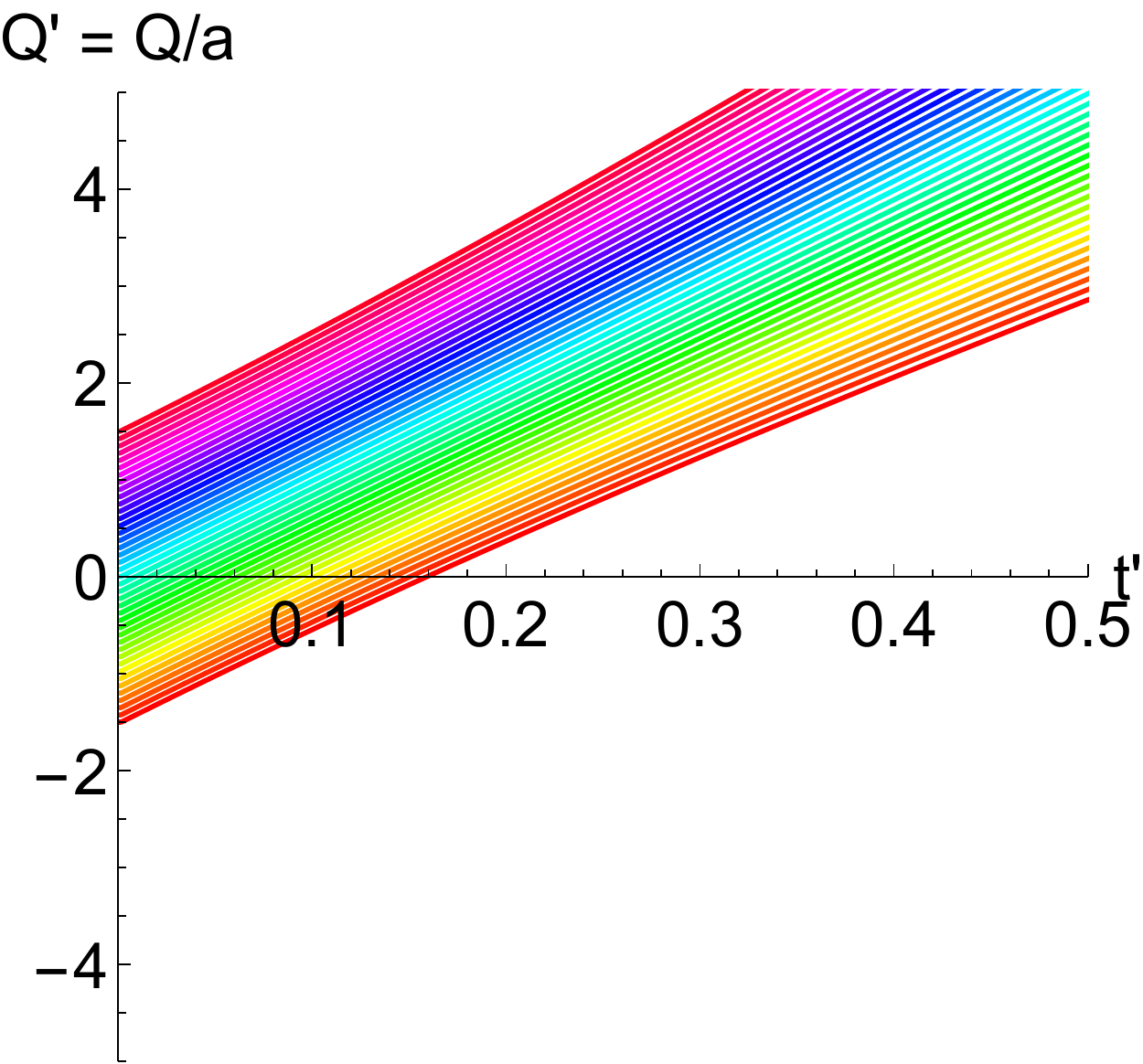}
\end{minipage}
\hfill
\begin{minipage}[c]{0.32\textwidth}
\includegraphics[width=4cm]{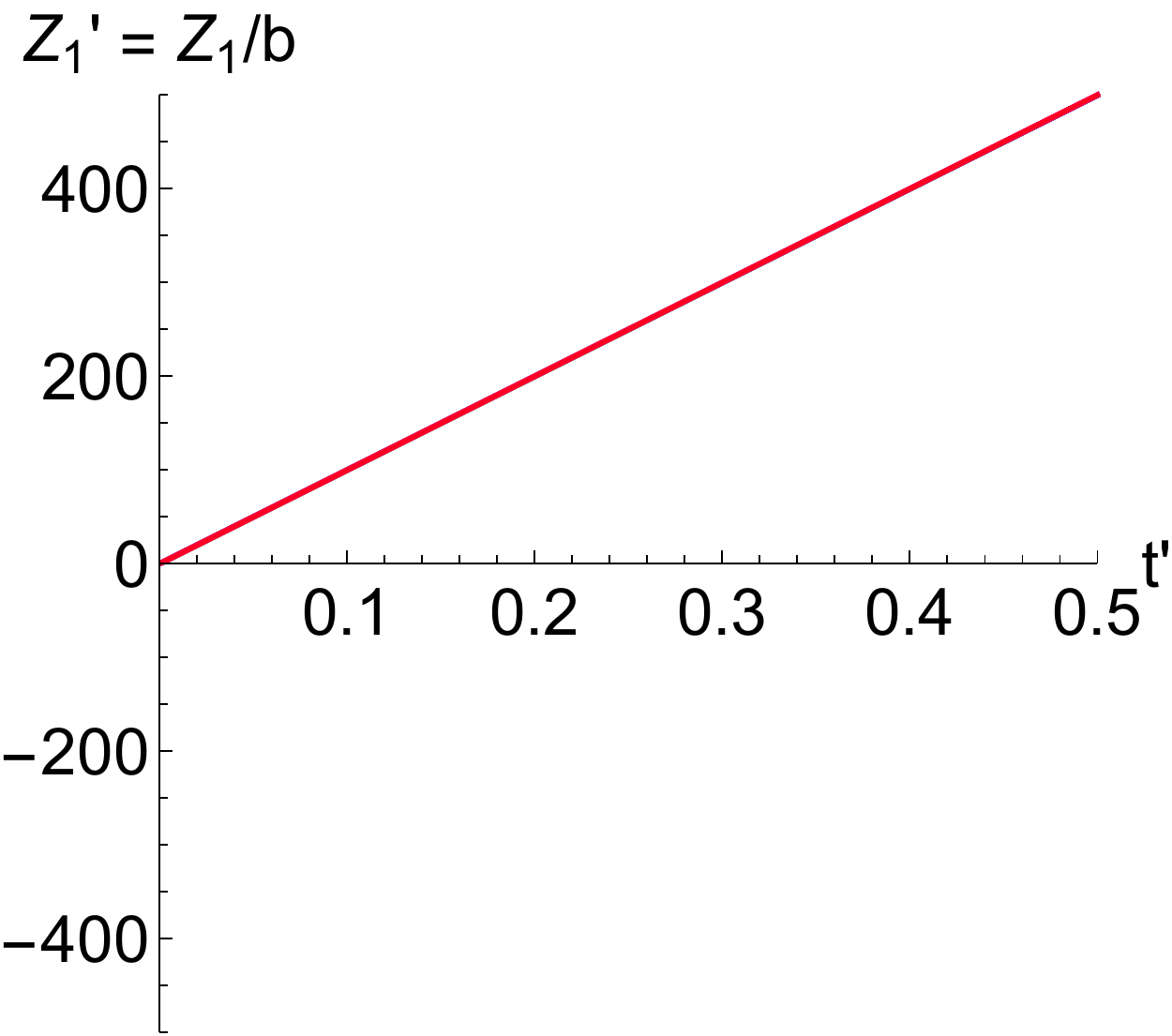}
\end{minipage}
\hfill
\begin{minipage}[c]{0.32\textwidth}
\includegraphics[width=4cm]{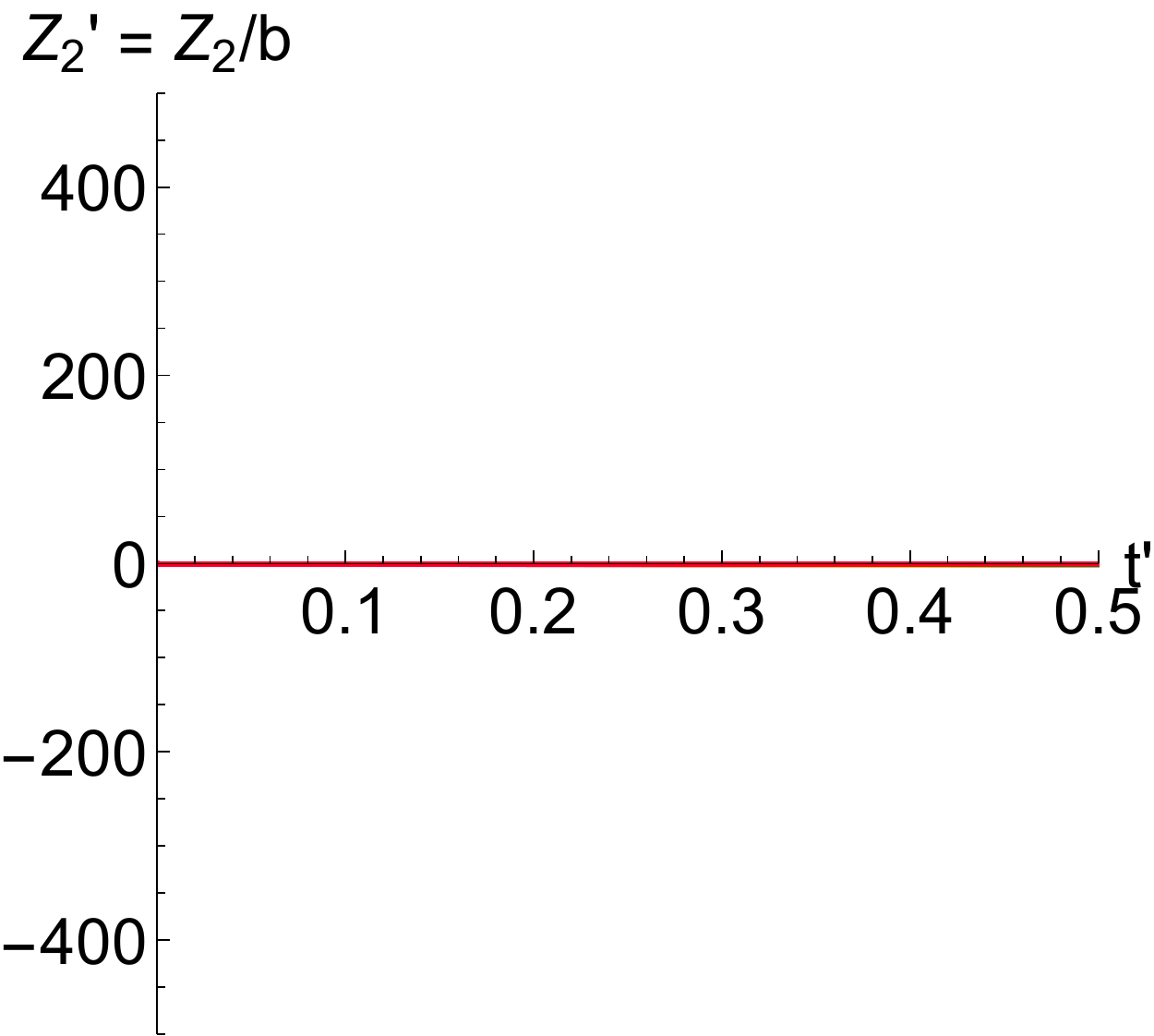}
\end{minipage}
\bigskip
\caption{Same situation as in figures~\ref{fig-3} and \ref{fig-5}  ($\sigma = 0$), but with $N=10^4$ pointers and smaller offset values for the initial positions of the pointers ($\hat{Z}_1 (0)=\hat{Z}_2 (0) =0.02$). Even with these small values, the initial positions of the particles in the pointers now completely determine the direction in which the spin particle flies, and therefore the final spin direction. The initial position of the spin particle plays no role in the determination of the result of measurement.}
\label{fig-6}
\vspace{2mm}
\hrule
\end{figure}

Figure~\ref{fig-6} is obtained in the same conditions as figure~\ref{fig-5}, but with a much larger number of pointer particles ($N=10^4$ ) and only a very small positive offset of the initial Bohmian positions $\hat{Z}_{1,2} (0) = 0.02$. A striking effect is that this small offset is sufficient to force all trajectories of the spin particle to fly upwards at the output of the measurement apparatus, whatever the initial position of this particle. In this case,  we see that it is really the pointers that \textquotedblleft decide\textquotedblright\ what the measurement result should be and, so to say, force the particle to \textquotedblleft obey to this decision\textquotedblright\ and to take a spin up value with a trajectory flying upwards.

 This \textquotedblleft all or nothing\textquotedblright\ effect of the pointer position shows very clearly that, what determines the final result of measurement is not the Bohmian position of the spin particle, but rather those of the particles inside the measurement apparatus. The standard quantum results are nevertheless recovered when an average is taken over all initial positions of the pointers, as we have checked numerically.

\subsection{Nonzero initial spin polarization}

When the initial spin polarization does not vanish, it introduces a preferred result of measurement, and therefore a preferred deviation of the particle trajectory. In the limiting case where the spin is fully polarized in one direction, one of the two components of the many body wave function (\ref{2}) vanishes, and the final direction of motion of all the Bohmian positions is fixed; the initial value of the position of the spin particle plays no role in the result of measurement.

It is more interesting to study intermediate situations, where a compromise has to occur between the quantum mechanical preference for one of the results and the influence of the initial values of the Bohmian positions.

\subsubsection{Few particles in the pointers}

\begin{figure}[t]  
\centering
\begin{minipage}[c]{0.32\textwidth}
\includegraphics[width=4cm]{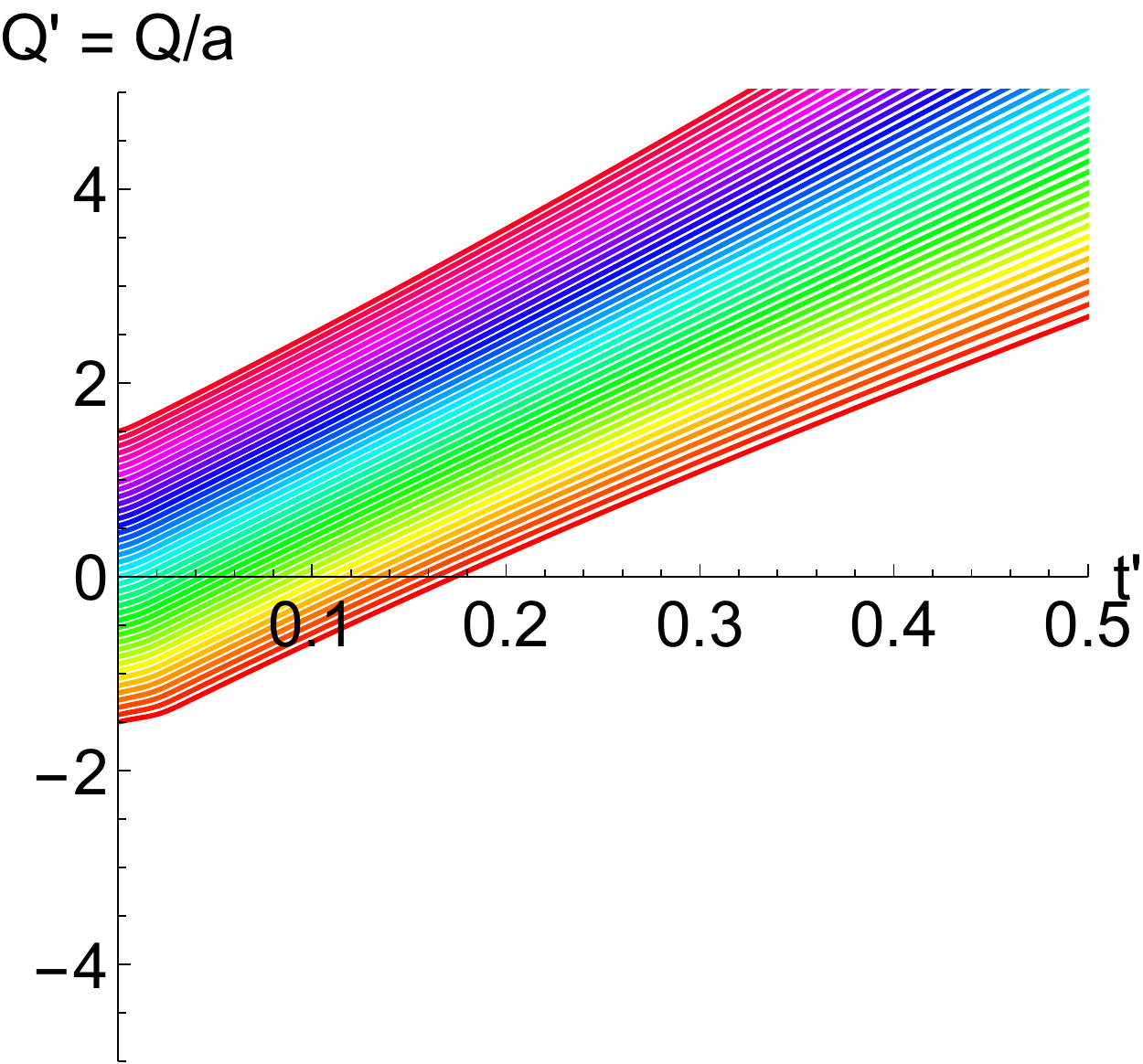}
\end{minipage}
\hfill
\begin{minipage}[c]{0.32\textwidth}
\includegraphics[width=4cm]{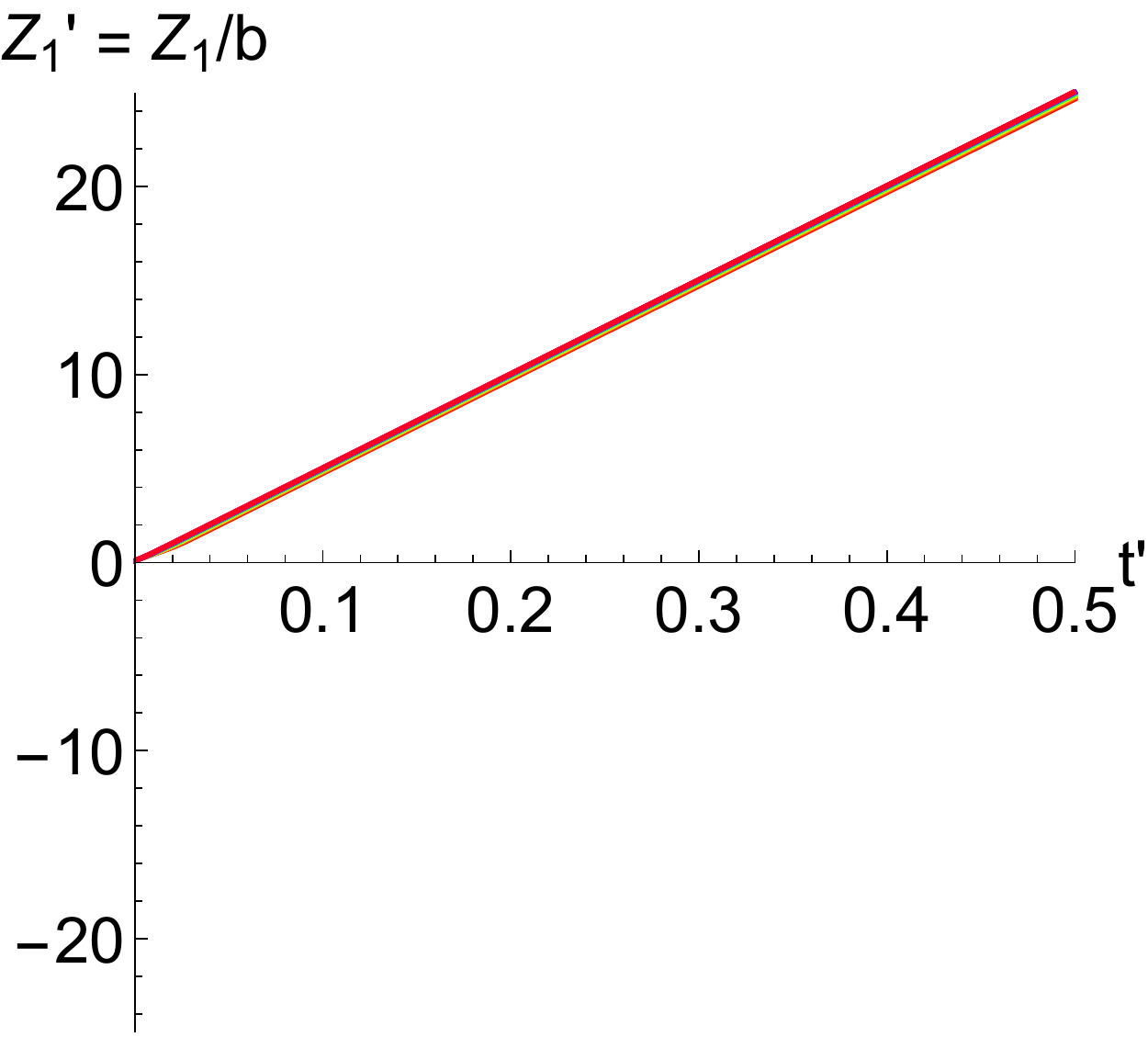}
\end{minipage}
\hfill
\begin{minipage}[c]{0.32\textwidth}
\includegraphics[width=4cm]{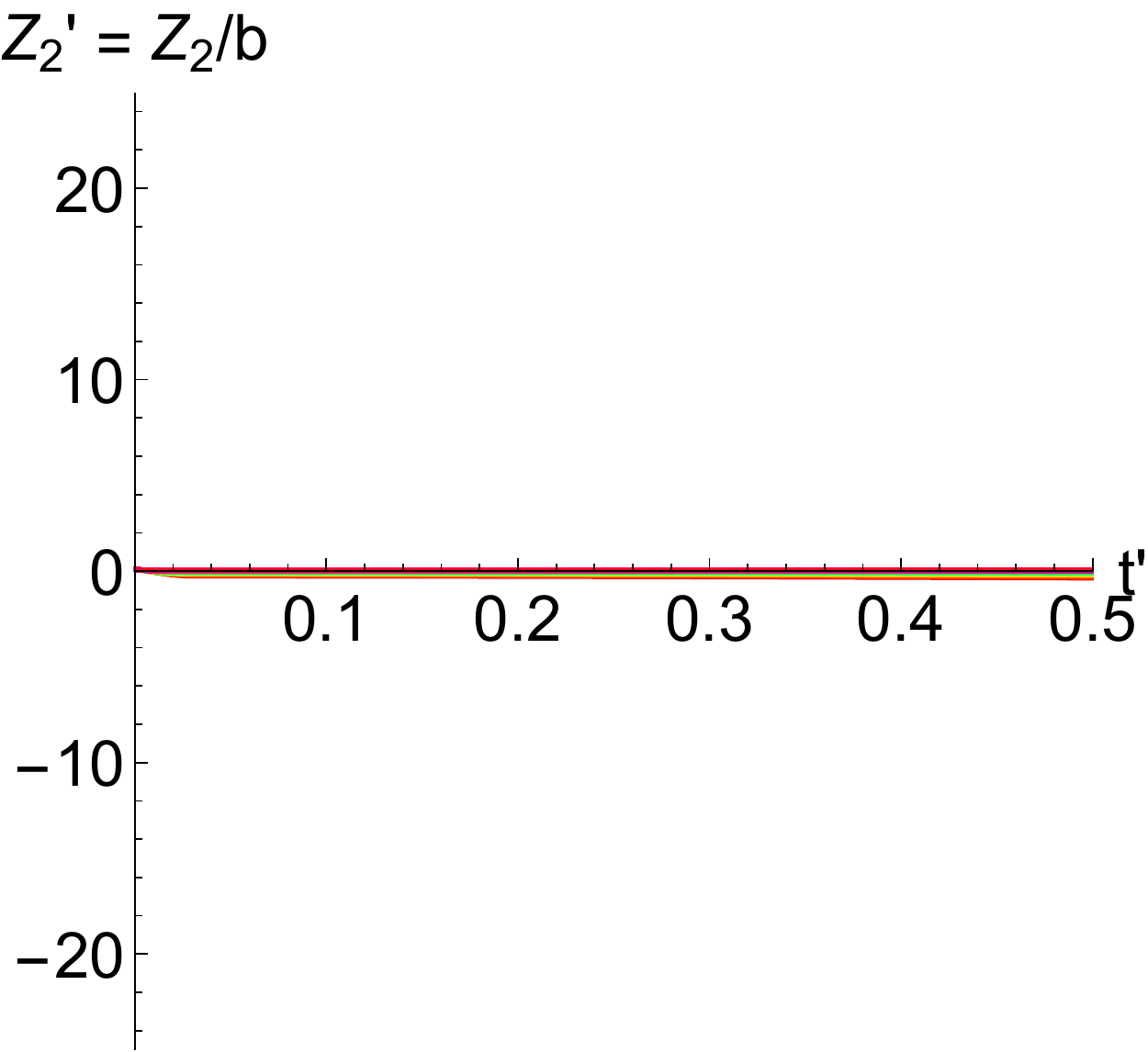}
\end{minipage}
 \bigskip

\caption{Bohmian trajectories of the spin particle and the particles of the pointers in a case where the initial polarization of the spin is $\sigma = 0.5$. Each pointer contains 25 particles. If the Bohmian positions of the pointers favor a spin up result ($\hat{Z}_1 (0)=\hat{Z}_2 (0)= 0.1$), the effect of the pointers is to force all the spin particle to go upwards.}
\vspace{2mm}
\hrule
\label{fig-7}
\end{figure}

\begin{figure}[!b]  
\centering
\hrule
\vspace{2mm}
\begin{minipage}[c]{0.32\textwidth}
\includegraphics[width=4cm]{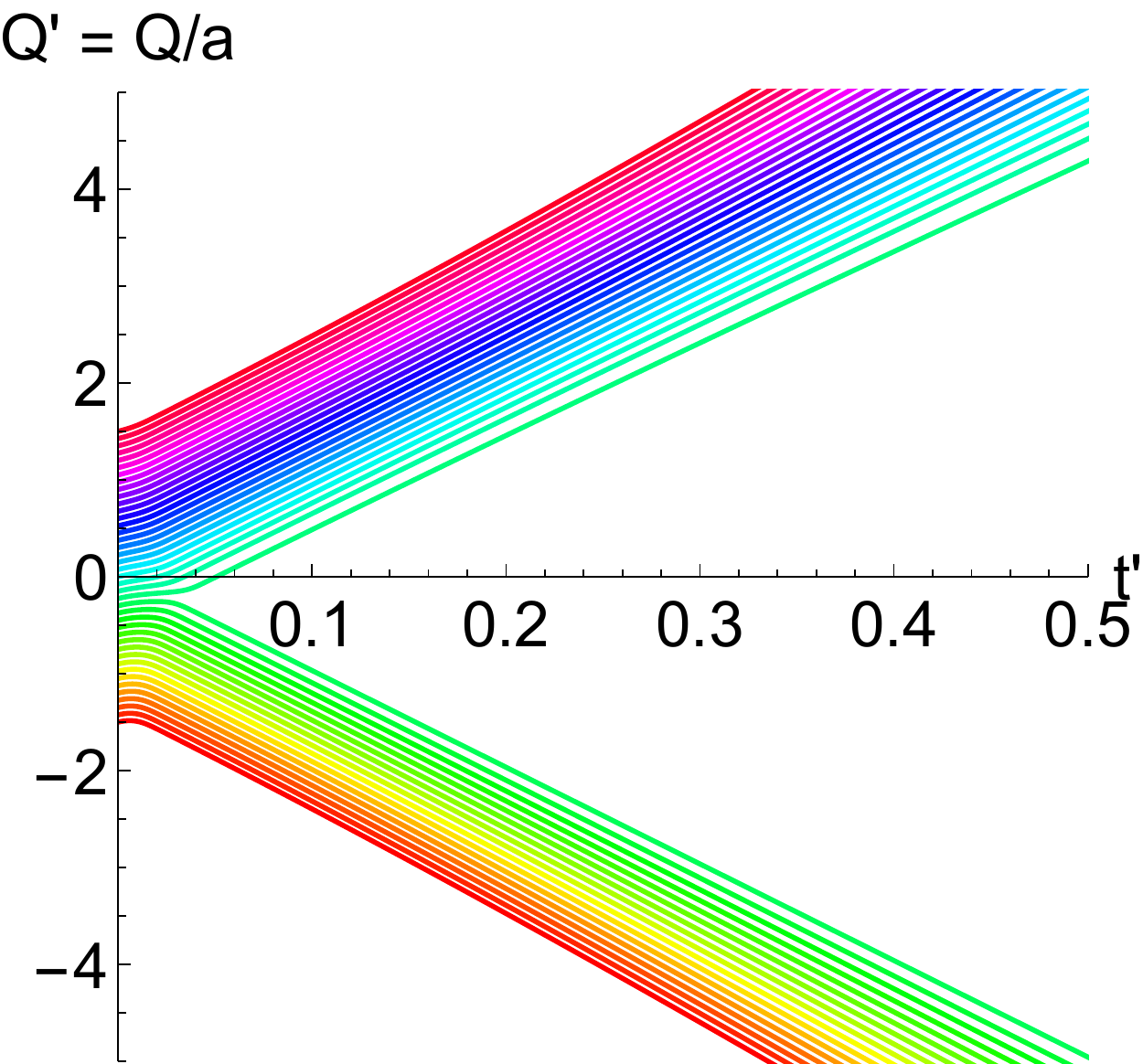}
\end{minipage}
\hfill
\begin{minipage}[c]{0.32\textwidth}
\includegraphics[width=4cm]{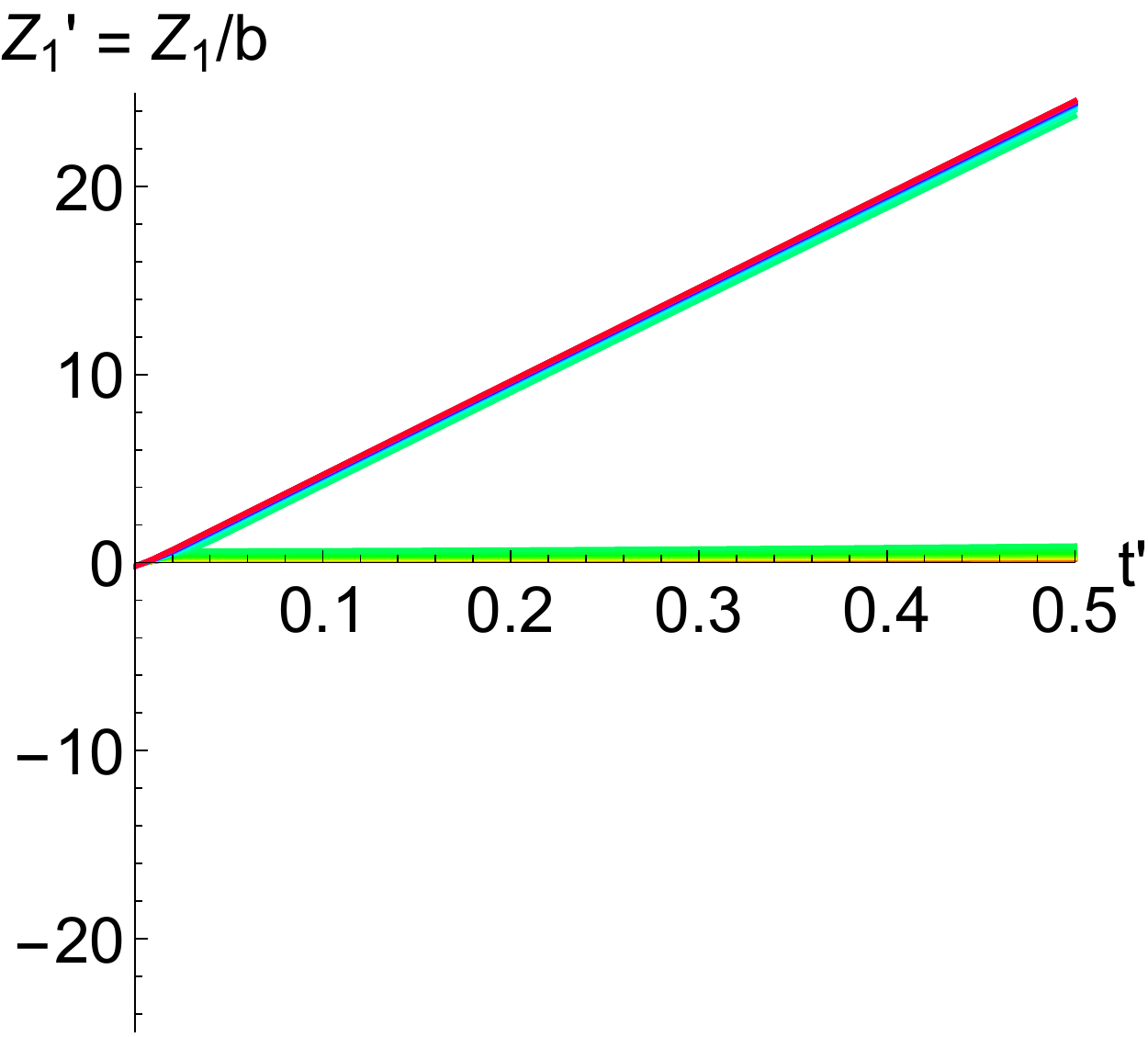}
\end{minipage}
\hfill
\begin{minipage}[c]{0.32\textwidth}
\includegraphics[width=4cm]{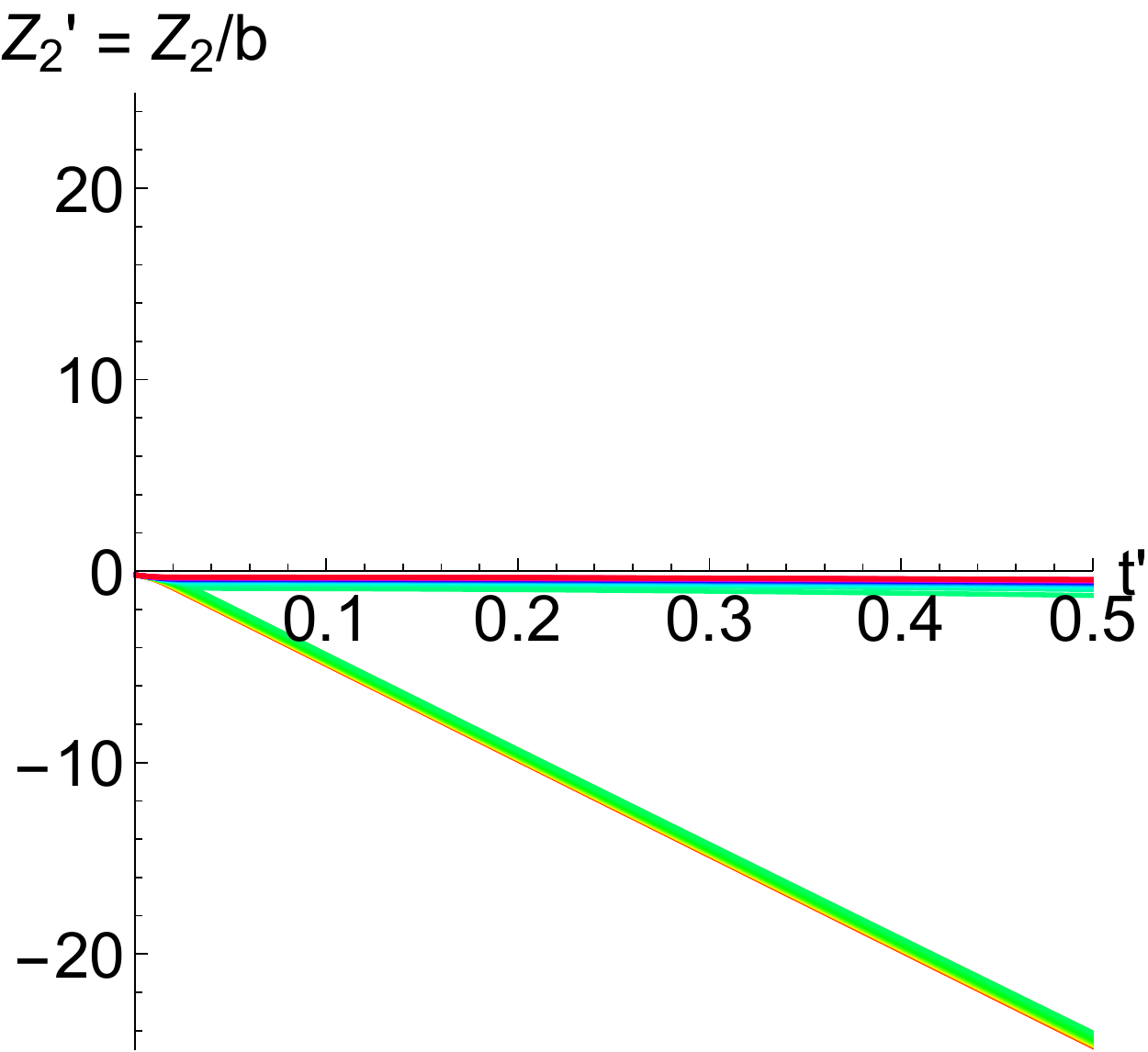}
\end{minipage}
 \bigskip
\caption{As in figure \ref{fig-7}, the initial spin polarization is $\sigma = 0.5$, but the initial positions of the  $25$ pointer particles now favor negative results: $\hat{Z}_1 (0)=\hat{Z}_2 (0) = - 0.2$. A significant proportion of the trajectories of the spin particle now go downwards. An average over all possible values of $\hat {Z}_{1,2}$ with a weight equal to the quantum probabilities (quantum equilibrium) would show that $3/4$ of the trajectories fly upwards and $1/4$ downwards, which corresponds to the standard quantum result.}
\label{fig-8}
\end{figure}

Figure~\ref{fig-7} shows the trajectories when the spin polarization is $\sigma = + 0.5$, assuming that the average initial positions of the 25 pointer particles are positive. In this case, all particle trajectories move upwards, as if the spin polarization were 100\%. Figure~\ref{fig-8} shows what happens when the average initial positions of the pointers are negative. In this case, some trajectories go downwards. Again, the standard quantum average is recovered only when a statistical average over the initial positions of the pointer is applied.

\subsubsection{Many particles in the pointers}
\label{many-pointers-2}

\begin{figure}[t]  
\centering
\begin{minipage}[c]{0.32\textwidth}
\includegraphics[width=5cm]{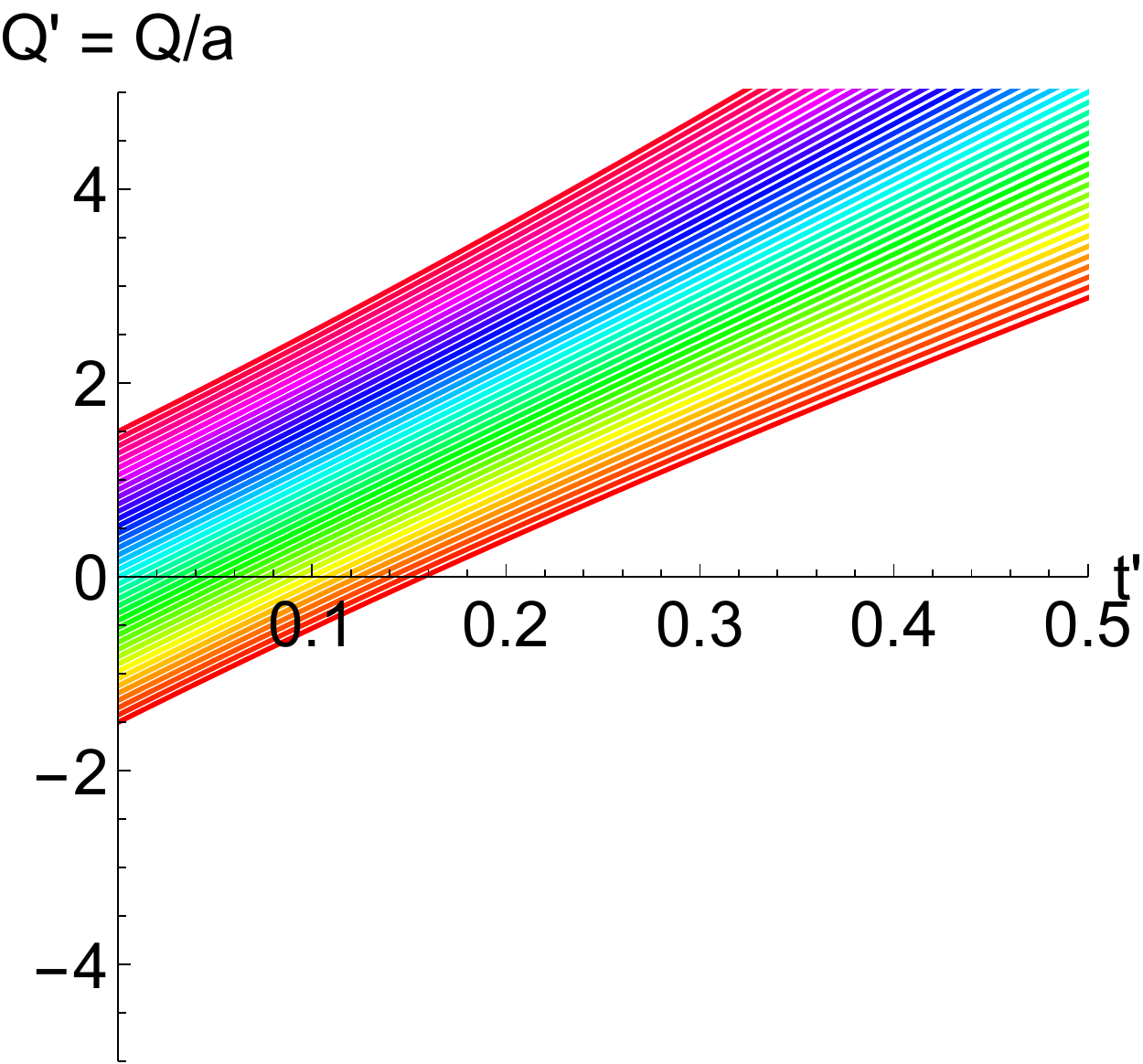}
\end{minipage}
\hfill
\begin{minipage}[c]{0.32\textwidth}
\includegraphics[width=5cm]{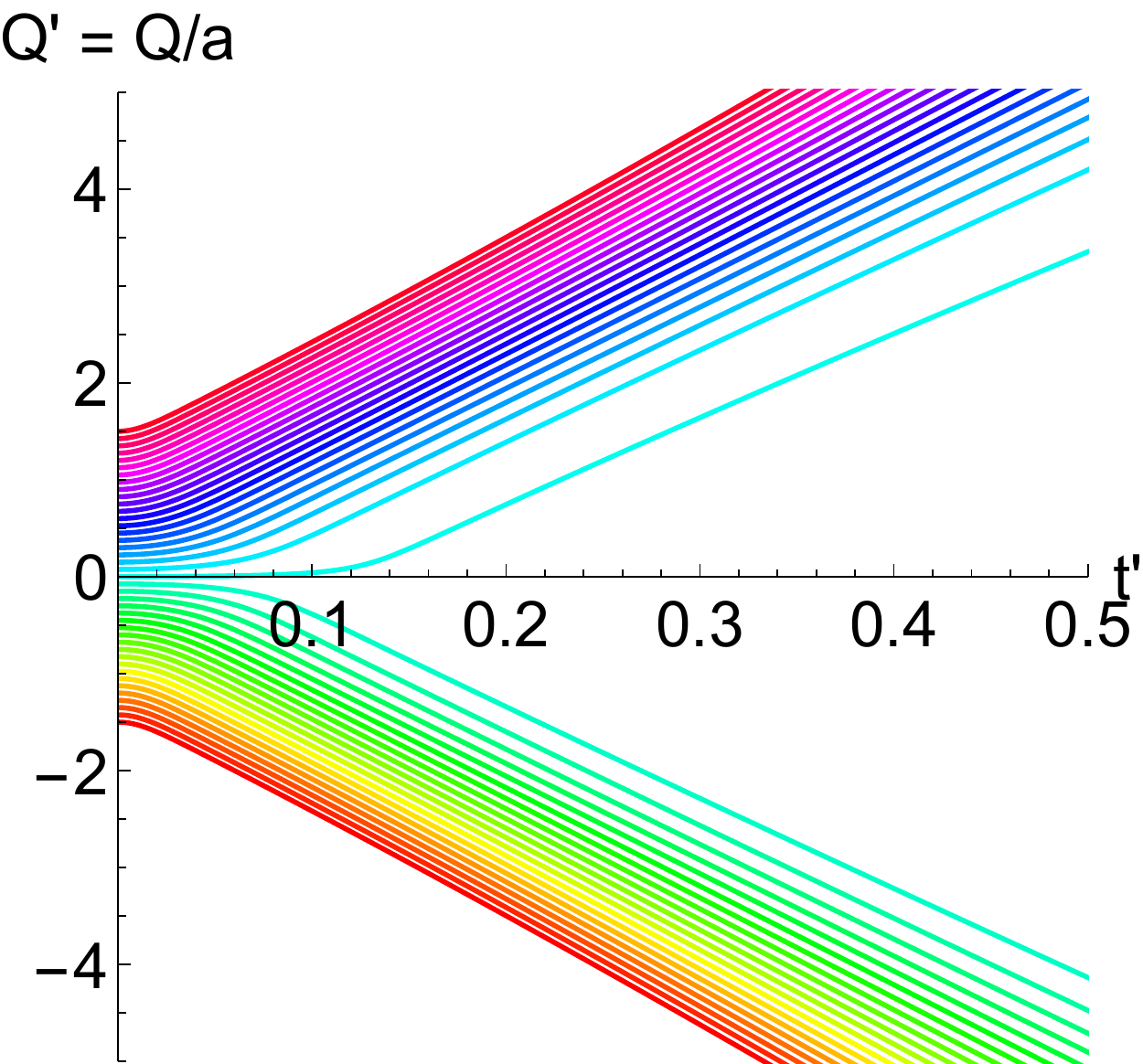}
\end{minipage}
 \bigskip
\caption{The left part of the figure shows the trajectories obtained when the initial spin polarization is positive but low, $\sigma = 0.01$; the number of the particles in the pointers is high, $N = 10^6$, while their initial positions are  neutral, $\hat{Z}_{1,2}(0)= 0 $. As a point of comparison, the right part shows the same situation, but in the absence of any entanglement with the pointers ($V'=0$); as expected, almost as many trajectories then fly up or down. This figure illustrates that, when the number of particles in the pointers is high, even if their initial positions are neutral, they completely change the trajectories of the spin particles, forcing  all of them to fly upwards and with a spin up. The initial Bohmian position of the spin particle is then completely irrelevant concerning the measurement result.}
\label{fig-9}
\vspace{2mm}
\hrule
\end{figure}

If the number of pointers is macroscopic, as in \S~\ref{many-pointers-1}, the Bohmian variables tend to completely dominate the determination of the result of measurement. The left part of figure~\ref{fig-9} shows the trajectories of the spin particle when the spin polarization  is very small, $\sigma = 0.01$, but $N$ is very large, $N=10^6$. The initial positions of the pointer particles are supposed to vanish ($\hat{Z}_{1,2}(0)= 0 $), so that they do not favour any result of measurement (they are \textquotedblleft neutral\textquotedblright ). In the absence of entanglement between the spin particle and the pointer particles, the numbers of trajectories going upwards and downwards should be almost equal, as illustrated in the right part of the figure (obtained by setting $V'=0$). One could then naively expect that neutral pointer particles cannot change this situation drastically. We see that this far from being true: actually, the Bohmian variables of the measurement apparatus can completely dominate the determination of the measurement result. They force all spin trajectories to remain grouped, in a sort of \textquotedblleft all or nothing\textquotedblright\ situation where all spins must take the same final positive value.
 The mechanism of this effect is as follows. Every particle in the first pointer has a slightly larger chance to have a positive initial velocity, and to move upwards. At subsequent times, its position corresponds to slightly larger values of $\left\vert \chi^{\pm}(Z_{p},t) \right\vert ^2 $ than $\left\vert \chi^{0}(Z_{p},t) \right\vert ^2 $, which favours the spin up  component of the total wave function; a similar effect takes place in the second pointer. Then  the multiplicative effect of $10^6$ particles transforms this small effect into a big unbalance, which affects the conditional wave function of the spin particle and makes it fly upwards.

\begin{figure}[t]  
\centering
\begin{minipage}[c]{0.32\textwidth}
\includegraphics[width=5cm]{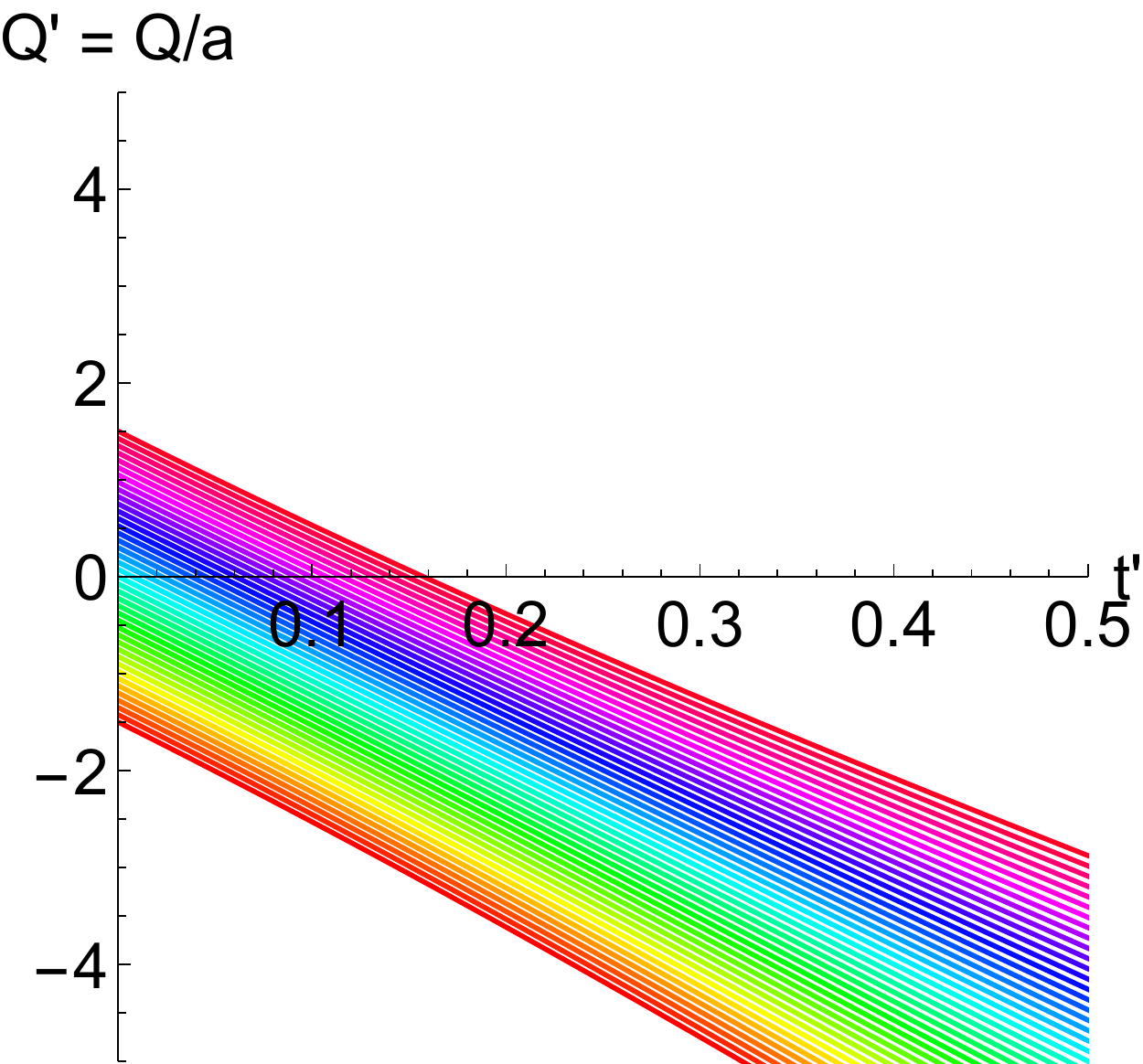}
\end{minipage}
\hfill
\begin{minipage}[c]{0.32\textwidth}
\includegraphics[width=5cm]{fig9-Q-no-coupling.pdf}
\end{minipage}
 \bigskip
\caption{Same conditions as in figure~\ref{fig-9} except that, here, the initial values of the positions of the center of mass of the pointer particles are slightly negative, corresponding to  $\hat{Z}_{1,2}(0)= -0.01 $. This is sufficient to completely change the  situation and reverse the deviation of all trajectories of the spin particle (left part of the figure; the right part is of course unchanged). Again, the initial Bohmian position of the spin particle is completely irrelevant.}
\label{fig-10}
\vspace{2mm}
\hrule
\end{figure}

Figure~\ref{fig-10} shows what happens when the initial values of the positions of the pointer particles are slightly negative, $\hat{Z}_{1,2}(0)= - 0.01 $. This very small change is sufficient to completely reverse the results of figure~\ref{fig-9}, since now all trajectories fly downwards. More generally, exploring many values of the parameters but keeping $N=10^6$ constant, we find that the trajectories of the spin particles (almost) always remain completely grouped, whatever the value of $\sigma$ is. This is of course in complete opposition with what happens in the absence of entanglement between S and M (right part of figures~\ref{fig-9} and \ref{fig-10}). Our main conclusion is that a large number of particles of the pointers renders the initial position of the spin particle completely irrelevant in the determination of the result of measurement. For each run of the experiment, this result is  pre-determined by the initial value of the positions of the particles inside the pointers.
\vspace{2mm}

\section{Discussion}

In their 1993 book, Bohm and Hiley \cite{Bohm-Hiley} emphasize that the
explicit context dependence of experimental results within dBB theory is
well in line with the central idea of the Copenhagen interpretation,
illustrated by a famous quotation by Bohr \cite{Bohr-1949}: \textquotedblleft
the necessity of considering the \textit{whole} experimental arrangement,
the specification of which is imperative for an well-defined application of
the quantum mechanical formalism\textquotedblright. Bohm and Hiley write:
\textquotedblleft The context dependence of measurements ... also embodies,
in a certain sense, Bohr's notion of the indivisibility of the combined
system of observing apparatus and observed object. Indeed, it can be said
that our approach (the dBB interpretation) provides a kind of intuitive
understanding of what Bohr was saying\textquotedblright.

Our discussion is fully in line with these remarks. It actually
gives them a more precise content, since it relates a general statement to a
specific mechanism involving all Bohmian positions variables in the measurement apparatus (context) and the measured system. They all interact through a common wave function, and the motion of the positions of the pointer particles  determines the final result of measurement\footnote{Needless to say, there is one exception: when  S is initially in an eigenstate of measurement, the result of measurement becomes certain, and the influence of all Bohmian variables (S and M) on the result disappears.}.
Our discussion also illustrates the effects of quantum
entanglement within dBB theory, where each particle position in S and M is driven by its own  \textquotedblleft
conditional wave function\textquotedblright\ \cite%
{conditional-wave-1,conditional-wave-2}, which depends on all other Bohmian positions. 
The question is then to determine whether the initial values of the
positions attached to S, or those attached to M, are dominant in the final state of the coupled motion. As we have seen, we have to distinguish between two cases, depending whether the pointers are fast or slow. 

In practice, fast pointers occur in many experiments, namely each time a big and fast amplification process  takes place within the measurement apparatus, which quickly entangles many particles of M with S. This is for instance the case if the apparatus includes photomultipliers, channeltrons, microchannel plates, drift chambers, etc. The many wave packets of the particles inside M then separate before the wave packets of S have the time to move, and the Bohmian positions of M  determine which wave packet drives the position of S. This results in a projection of the effective wave function of S, an effect that is very similar to state vector projection and to that discussed in Ref.\cite{GT-FL}; it is in general nonlocal.

Slow pointers occur in the opposite situation: when the wave packets of S separate before S gets entangled with many particles in M. In a standard Stern-Gerlach experiment, no amplification takes place, and this is probably the case. The atoms merely accumulate one by one on the output glass plate, long after their wave packets have completely separated in the field gradient. Of course, during the  initial interaction of the atom with the field gradient, a microscopic momentum is transferred to a large number $N$ of atoms or molecules, each of them receiving a recoil momentum proportional to $1/N$. We have seen that an amplification coefficient $\sqrt N$ appears in this case, but this is not sufficient to overcome the $1/N$ factor. As a consequence, we can expect that a Stern-Gerlach experiment corresponds to a case of slow pointers: therefore, the result of measurement is indeed determined by the initial position of the atom, as assumed in most discussions of this experiment.

Our conclusions also have implications concerning the calculation of correlation functions at different times. Within standard quantum mechanics, these functions are obtained by taking into account the first measurement  and the effect of the measurement apparatus. This is done by applying a projection operator, after which one calculates the subsequent evolution of the wave function from this new state. Similarly, in dBB quantum theory, it is essential to take into account the effect of the Bohmian variables of the measurement apparatus. Otherwise, incorrect correlation functions are obtained \cite{Morchio}.

Needless to say, the model of the measurement apparatus we have used is oversimplified, assuming for instance the same Gaussian wave function for all the particles inside the measurement apparatuses that are entangled with the spin particle. This simplification does not change the structure of the entangled quantum state of the spin particle and the measurement apparatus. Since this structure is the origin of our results, we believe that they are generic. Moreover, for simplicity, we have focussed our discussion on the particles inside the pointers, but it is clear that other parts of the measurement apparatuses also play a role. The number of particles $N$ should not be seen as referring only to the physical content of the pointers themselves. It should be understood as the number of particles that get entangled with the spin particle during the amplification process that takes place in any measurement apparatus and results in the physical displacement of the pointers.

\section{Conclusion}

Curioulsy, many discussions of quantum measurements within dBB theory emphasize the role of the Bohmian position(s) associated  with the microscopic system S,  ignoring all those attached to the measurement apparatus M. Under these conditions, of course, the result of a
measurement performed on S can reveal nothing but the
initial value of the Bohmian position(s) attached to S. Nevertheless, we have learnt from the historical discussions between Einstein and Bohr \cite{Bohr-1949}, for instance the argument concerning an interference experiment with a moving pierced screen playing the role of a which way apparatus, that the quantum properties of the measurement apparatus cannot be ignored without running into contradictions. Indeed, the same rules have to be applied to both  S and M; otherwise one misses an essential purpose of the dBB  theory, which is to propose a framework for a completely unified dynamics.

During the initial stage of a measurement process, S becomes entangled with some variables of M, so that the velocity of the Bohmian position variables attached to S depend on those of M.  Our analysis takes into account this entanglement and studies in detail how it changes the way the wave function of the whole system drives every position. It shows that the initial value of the positions attached to M play a crucial role. This also
 illustrates the intrinsic contextuality \cite%
{Bell-1966,Kochen-Specker,DWRUQM} of the dBB theory: the results associated
with a given observable do not depend only on some initial value of a
variables pertaining to S; actually, for a macroscopic measurement apparatus, we have seen that they may even be independent of the positions attached to S. This is in direct line with Bohr's ideas, where the results crucially depend on the entire
measurement apparatus which, in a Bohmian context, means that they depend on all the initial positions inside this apparatus. See also \cite{Nikolic} for a general discussion of why microscopic trajectories play only an auxiliary role in Bohmian mechanics formulated in terms of macroscopic phenomena.

Another way to describe the effect of entanglement is to emphasize its nonlocal aspect. When S no longer interacts with M, the wave packets of the pointers may still overlap, so that no specific result of measurement has emerged yet. Later, when the wave packets of the pointers have significantly moved, this overlap vanishes, forcing the ensemble of pointer positions to \textquotedblleft choose\textquotedblright\ a component of the wave function where the result is determined. In such a case, it is a nonlocal effect originating from the measurement apparatus that drives S into the wave packet associated with the result of measurement. The microscopic system has, so to say, to \textquotedblleft follow the decision\textquotedblright\ of the macroscopic system. The left parts of figures~\ref{fig-9} and \ref{fig-10} illustrate how the initial position of the spin particle then plays no role whatsoever in the determination of the measurement result.  This can be seen as the macroscopic counterpart of the nonlocal effect arising in a Bell experiment performed with two spins and two Stern-Gerlach apparatuses, where the position of one microscopic particle drives the position of the other particle.


\end{document}